\documentclass[11pt,a4paper]{article}

\usepackage{amsmath,amsfonts,amssymb,amsthm,cite}
\usepackage{tikz} 

\date{30 January 2008}

\title{Fourier analysis on the affine group, \\
quantization and noncompact Connes geometries}

\author{Victor Gayral,$^1$ \ Jos\'e M. Gracia-Bond\'{\i}a$^{2,3}$
\ and \ Joseph C. V\'arilly,$^4$
\\[1pc]
$^1$\,Laboratoire de Math\'ematique,
Universit\'e de Reims Champagne--Ardenne, \\
51687 Reims, France
\\[6pt]
$^2$\,Departamento de F\'{\i}sica Te\'orica,
Universidad de Zaragoza,
Zaragoza 50009, Spain
\\[6pt]
$^3$\,Departamento de F\'{\i}sica, Universidad de Costa Rica,
2060 San Jos\'e, Costa Rica
\\[6pt]
$^4$\,Departamento de Matem\'aticas, Universidad de Costa Rica,
\\
2060 San Jos\'e, Costa Rica}

\numberwithin{equation}{section}

\newtheorem{thm}{Theorem}[section]
\newtheorem{lem}[thm]{Lemma}
\newtheorem{corl}[thm]{Corollary}
\newtheorem{prop}[thm]{Proposition}

\theoremstyle{definition}

\theoremstyle{remark}

\DeclareMathOperator{\ad}{ad}           
\DeclareMathOperator{\Ad}{Ad}           
\DeclareMathOperator{\Aff}{Aff}         
\DeclareMathOperator{\Coad}{Coad}       
\DeclareMathOperator{\Dom}{Dom}         
\DeclareMathOperator{\Gal}{Gal}         
\DeclareMathOperator{\HS}{HS}           
\DeclareMathOperator{\Ind}{Ind}         
\DeclareMathOperator{\sinch}{sinch}     
\DeclareMathOperator{\Tr}{Tr}           
\DeclareMathOperator{\tr}{tr}           

\newcommand{\al}{\alpha}                
\newcommand{\bt}{\beta}                 
\newcommand{\Dl}{\Delta}                
\newcommand{\dl}{\delta}                
\newcommand{\Ga}{\Gamma}                
\newcommand{\ga}{\gamma}                
\newcommand{\ka}{\kappa}                
\newcommand{\La}{\Lambda}               
\newcommand{\la}{\lambda}               
\newcommand{\nb}{\nabla}                
\newcommand{\Om}{\Omega}                
\newcommand{\om}{\omega}                
\newcommand{\sg}{\sigma}                
\renewcommand{\th}{\theta}              
\newcommand{\vf}{\varphi}               

\newcommand{\A}{\mathcal{A}}            
\newcommand{\aff}{\mathfrak{aff}}       
\newcommand{\B}{\mathcal{B}}            
\newcommand{\D}{\mathcal{D}}            
\renewcommand{\d}{\mathbb{d}}           
\newcommand{\downto}{\downarrow}        
\newcommand{\Coo}{C^\infty}             
\newcommand{\del}{\partial}             
\newcommand{\delbar}{\bar\partial}      
\newcommand{\Dslash}{{D\mkern-11.5mu/\,}} 
\newcommand{\Daslash}{{D\mkern-9.5mu\backslash\,}} 
\newcommand{\EE}{\mathbb{E}}            
\newcommand{\EEmod}{\mathbb{E}^{\mathrm{mod}}} 
\newcommand{\FAFK}{\mathbb{F}_{\scriptscriptstyle\!AFK}} 
\newcommand{\FK}{\mathbb{F}_{\scriptscriptstyle\!K}} 
\newcommand{\FM}{\mathbb{F}_{\scriptscriptstyle\!M}} 
\newcommand{\FMmod}{\FM^{\mathrm{mod}}} 
\newcommand{\g}{\mathfrak{g}}           
\newcommand{\gsl}{\mathfrak{sl}}        
\renewcommand{\H}{\mathcal{H}}          
\newcommand{\h}{\mathfrak{h}}           
\newcommand{\half}{{\mathchoice{\thalf}{\thalf}{\shalf}{\shalf}}}
\newcommand{\hatox}{\mathrel{\widehat\otimes}} 
\newcommand{\hideqed}{\renewcommand{\qed}{}} 
\newcommand{\K}{\mathcal{K}}            
\newcommand{\lt}{\triangleright}        
\newcommand{\Oh}{\mathcal{O}}           
\newcommand{\otto}{\leftrightarrow}     
\newcommand{\ottto}{\longleftrightarrow} 
\newcommand{\owl}{\overline}            
\renewcommand{\P}{\mathcal{P}}          
\newcommand{\PB}{\mathrm{PB}}           
\newcommand{\quarter}{\tfrac{1}{4}}     
\newcommand{\R}{\mathbb{R}}             
\newcommand{\rt}{\triangleleft}         
\newcommand{\shalf}{{\scriptstyle\frac{1}{2}}} 
\newcommand{\sym}{\mathrm{sym}}         
\newcommand{\thalf}{\tfrac{1}{2}}       
\newcommand{\tihalf}{\tfrac{i}{2}}      
\newcommand{\tribar}{|\mkern-2mu|\mkern-2mu|} 
\newcommand{\U}{\mathcal{U}}            
\newcommand{\w}{\wedge}                 
\newcommand{\x}{\times}                 
\newcommand{\7}{\dagger}                
\newcommand{\8}{\bullet}                
\renewcommand{\.}{\cdot}                
\renewcommand{\:}{\colon}               

\newcommand{\bilin}[2]{\langle#1,#2\rangle} 

\newcommand{\braket}[2]{\langle#1\mathbin|#2\rangle} 
\newcommand{\ketbra}[2]{\lvert#1\rangle\langle#2\rvert} 
\newcommand{\pd}[2]{\frac{\partial#1}{\partial#2}} 
\newcommand{\set}[1]{\{\,#1\,\}}        
\newcommand{\snorm}[1]{\mathopen{\tribar}{#1}\mathclose{\tribar}}
\newcommand{\Twobytwo}[4]{\begin{pmatrix}#1 & #2 \\[\jot] #3 & #4
\end{pmatrix}} 
\newcommand{\twobytwo}[4]{\begin{pmatrix}#1 & #2 \\ #3 & #4
\end{pmatrix}} 
\newcommand{\word}[1]{\quad\mbox{#1}\quad} 
\def\<#1,#2>{\langle#1\mathbin|#2\rangle} 

\makeatletter
\def\section{\@startsection{section}{1}{\z@}{-3.5ex plus -1ex minus
			  -.2ex}{2.3ex plus .2ex}{\large\bf}}
\def\subsection{\@startsection{subsection}{2}{\z@}{-3.25ex plus -1ex
			  minus -.2ex}{1.5ex plus .2ex}{\normalsize\bf}}
\renewcommand{\@dotsep}{200} 
\makeatother

\begin{document}

\maketitle

\begin{abstract}
We find the Stratonovich--Weyl quantizer for the nonunimodular affine
group of the line. A noncommutative product of functions on the
half-plane, underlying a noncompact spectral triple in the sense of
Connes, is obtained from it. The corresponding Wigner functions
reproduce the time-frequency distributions of signal processing. The
same construction leads to scalar Fourier transformations on the
affine group, simplifying and extending the Fourier transformation
proposed by Kirillov.
\end{abstract}



\begin{flushright}
\textit{Dedicated to Orietta Protti, who started it all}
\end{flushright}

\section{Introduction}

The theory of \textit{noncompact spin geometries} in the sense of
Connes~\cite{ConnesCollege,ConnesGrav} or \textit{noncompact spectral
triples}, broached in~\cite{Selene}, was developed in references
\cite{RennieSmooth,Himalia,Massalia}. The coordinate algebras treated
in~\cite{RennieSmooth} have locality properties analogous to those of
the commutative case, whereas \cite{Himalia,Massalia} deal with truly
noncommutative contexts that are essentially \textit{flat},
respectively the Moyal $2n$-planes and some of their generalizations.

As such, the theory remains underdeveloped. This is partly for want of
suitable noncommutative noncompact spectral triples. Among the myriads
of deformations or ``star products'', twisted product algebras with
the crucial \textit{traciality} property distinguish themselves in
that the classical integral yields a faithful tracial state. This is
what made the original (Groenewold--)\allowbreak Moyal product
\cite{Groenewold,Moyal,BayenFFLS} so popular in quantum field theory
\cite{Filk,Atlas,SeibergW}. Not least, it ensures relevant properties
in cyclic cohomology~\cite{ConnesFS}.

Let~$X$ be a phase space, $\mu$ a convenient measure on it (often the
Liouville measure) and~$\H$ the Hilbert space associated to $(X,\mu)$.%
\footnote{This paragraph and the next are excerpted from the report by
one of us (JMGB) to the Oberwolfach conference on Dirac Operators and
Noncommutative Geometry, in November 2006.}
Denote by $\dl_\mu(x,x')$ the reproducing kernel for~$\mu$. A
\textit{Stratonovich--Weyl quantizer} or tracial quantizer for
$(X,\mu,\H)$ is an operator-valued distribution $\Om$ on~$X$, with
values in the space of selfadjoint operators on~$\H$, spanning a
weakly dense subset of $\B(\H)$, and verifying
\begin{equation*}
\Tr\Om(x) = 1, \qquad \Tr\bigl[\Om(x)\Om(x')\bigr] = \dl_\mu(x,x').
\end{equation*}
Quantizers in this sense, if they exist, are essentially unique.
Ownership of a quantizer solves in principle all quantization
problems: \textit{quantization} of a (sufficiently regular) function
or ``symbol'' $a$ on~$X$ is effected by
$$
a \mapsto \int_X a(x) \Om(x)\,d\mu(x) =: Q(a),
$$
and dequantization of an operator $A \in B(\H)$ is achieved by
\begin{equation}
A \mapsto \Tr\bigl[ \Om(\.) A \bigr] =: W_A(\.).
\label{eq:mouse-trap}
\end{equation}
Indeed $\Om$ can just as well be called a \textit{dequantizer}. It
follows that $1_{\H}\mapsto1$ by dequantization, and also
$$
\Tr Q(a) = \int_X a(x) \,d\mu(x).
$$
Moreover, since the set $\Om(X)$ is total, it is clear that
$$
W_{Q(a)}(x) = \Tr\biggl[ \biggl(\int_X a(x')\,\Om(x')\,d\mu(x')
\biggr) \Om(x) \biggr] = a(x),
$$
so $Q$ and~$W$ are inverse to one another. In particular,
$W_{Q(1)} = 1$ says that $1 \mapsto 1_{\H}$ by quantization. Finally,
the following relation holds:
\begin{equation}
\Tr\bigl[ Q(a) Q(b) \bigr] = \int_X a(x) b(x) \,d\mu(x).
\label{eq:alpha-omega}
\end{equation}
This is the tracial property.

Most of the interesting cases occur in the context of group actions;
that is to say, there is a Lie group~$G$ for which $X$ is a symplectic
homogeneous $G$-space, with $\mu$ then being a $G$-relatively
invariant measure on~$X$, and $G$ acts by a (multiplier) unitary
irreducible representation $U$ on~$\H$. A quantizer for the data set
$(X,\mu,\H,G,U)$ satisfies the previous defining equations and is
endowed with the covariance property:
$$
U(g) \,\Om(x)\, U^\7(g) = \Om(g \lt x),
$$
for all $g \in G$, $x \in X$. Orbits of the \textit{coadjoint} action
of $G$ on its Lie algebra dual $\g^*$ and symplectic homogeneous
manifolds are essentially the same thing~\cite[Sect.~1.4]{Kirillov},
and in this paper we think of the action denoted by~$\lt$ above as an
instance of the coadjoint action of~$G$.

A covariant collection as above, but satisfying only
$$
\int_X \Om(x)\,d\mu(x) = 1_{\H} \word{and} \Tr\Om(x) = 1,
$$
may be called a semitracial quantizer.

\vspace{6pt}

Once in possession of the quantizer, one can in principle immediately
construct a twisted product that will be normalized (its identity
being the constant function~$1$), hermitian (complex conjugation being
the involution), covariant under an appropriate group action, and
tracial. We give the details in the body of the paper.

In summary, the ``Stratonovich--Weyl'' label here refers neither to
general deformations nor to star products obtained (roughly speaking)
by reduction~\cite{Selene}, extension~\cite{Melpomene} or induction
from the original one; but to a restricted category, defined by a
precise set of postulates, designed to capture the main trait behind
the success of Moyal's formalism for Quantum Mechanics; namely, that
quantum and classical expected values should be computed by the same
rule. The products thereby obtained are non-formal and analytically
controlled. The quest for quantizers in our sense is richly rewarding.
In this paper we show by example how the theory of covariant tracial
quantizers meshes with, and substantially complements, Kirillov's
method of orbits in representation theory. Its main end products are
the (scalar) Fourier--Moyal kernels on $\g^* \x G$:
$$
\EE(x,g)    := \Tr\bigl[ \Om(x)\,U(g)          \bigr];  \qquad
\EEmod(x,g) := \Tr\bigl[ \Om(x)\,U(g)\sqrt{d}\,\bigr],
$$ 
with $d$ the formal dimension operator for~$U$. There is a good case,
that we have made before~\cite{Dione} and we make here again, for
these to be the central objects in harmonic analysis. For, as soon as
a tracial quantizer is available, the abstract Plancherel theorem of
group Fourier transform theory becomes a concrete one on the coadjoint
orbits. Also, quantizers have an important applied side, with their
relations to wavelets in signal processing and to quantum optics.

Before giving the usual guide to the article, we briefly return to
noncompact spectral triples. It seems natural to look for
noncommutative tracial algebras on the surface of constant negative
curvature, whether it be modelled by a hyperboloid in $\R^3$, the unit
disk, or the Poincar\'e upper half-plane~$\Pi$; we shall focus on the
half-plane. Even so, the prescription that the noncommutative
coordinate algebras on~$\Pi$ carry the full $SL(2,\R)$ symmetry may be
too much to ask, because then the first-order condition
of~\cite{ConnesCollege,ConnesGrav} for the standard Dirac operator
on~$\Pi$ cannot be satisfied. This will be shown in
Section~\ref{sec:geometry}. One can of course inherit the full
symmetry and use the alternative Dirac operator; or perhaps
``deform'' suitably the latter.%
\footnote{We thank Pierre Bieliavsky for illuminating discussions
of these aspects.}

Meanwhile we concentrate on the smaller `$ax + b$'-type group of
symmetry of~$\Pi$. As mentioned, there is another compelling
motivation for revisiting the Stratonovich--Weyl quantizers: the
progress of photonics~\cite{MarquardtEtal} allows nowadays a
quasi-measurement of the Wigner functions ---see
\cite{Royer,VogelR,BanaszekW} as theoretical harbingers. This is
calling for investigation of systems with solvable group symmetry; and
before tackling them, it helps to visit the case of affine-group
symmetry.

\vspace{6pt}

In the next section, we remind the reader of the Kirillov method for
constructing the unitary irreducible representations (unirreps) of the
`$ax + b$' group, in order to make the exposition self-contained. We
exhibit the corresponding characters and the Duflo--Moore operator for
the nontrivial representations. Section~3 is a modicum of real
analysis, eventually needed for establishing the properties of the
affine-group quantizers.

Sections 4 to~8 deal with the main issue. As it happens, the
literature on wavelets and time-frequency distributions already
contains the information needed to extract the quantizers on the
half-plane~\cite{BertrandsAff,BertrandsSymb}: an impressive example
that concrete problem-oriented work can lead to far-reaching
conceptual results. A suitable modification of the classical Weyl
quantization rule holds. We give the quantizers explicitly and verify
the key tracial property in Section~4. In the next section, we find
the (left-covariant) twisted product associated to the quantizer,
required for spectral triple theory, and investigate its symmetry
properties. Section~6 deals with the right-covariant counterparts for
the quantizer and the twisted product. In Section~7, the foregoing
illuminates harmonic analysis: \textit{scalar} Fourier--Moyal
transformations are found for the `$ax + b$' group, allowing us to
recover the representation characters and to improve on Kirillov's
Fourier transformation. The basic results of Fourier analysis, up to
and including the Plancherel formula, are shown to hold in our
context, for this nonunimodular case. Section~8 connects and compares
our formulation with others, including Fronsdal's
$\star$-representation program~\cite{Fronsdal} and the approach of a
remarkable series of recent papers~\cite{AliEtal,AliFK} by Ali and
coworkers, also inspired by the literature on wavelet transforms. We
take the occasion to set the record straight on the matter of
Stratonovich--Weyl quantizers.

Section 9 revisits the noncompact spectral triples over the half-plane
that have motivated the present work; the first-order property for the
Dirac operator can now be established. Section~10 gives pointers for
further \textit{rapprochement} of Connes', Kirillov's and Moyal's
paradigms \textit{inter alia}.

It remains to add that the standards of formality in this paper are
about the usual ones in mathematical physics; this saves considerable
spacetime. All our arguments are in fact rigorous, as will be shown
in~\cite{Eris}.

\section{The orbit method for the group of affine transformations}
\label{sec:Kirillov}

\subsection{The coadjoint orbits}
\label{ssc:coad}

The group~$\Aff$ of orientation-preserving linear transformations of
the real line, or affine group for short, also known as the `$ax + b$'
group, is a semidirect product $\R_+^\x \rtimes \R$, with the handy
matrix realization
\begin{equation}
\Aff \equiv \biggl\{\twobytwo{a}{b}{0}{1} : a > 0,\ b \in \R \biggr\}.
\label{eq:affine-group}
\end{equation}
Its Lie algebra is realized by the matrices
$$
\aff := \biggl\{X = \twobytwo{u}{v}{0}{0} : (u,v) \in \R^2 \biggr\},
\qquad  X = u X_1 + v X_2,
$$
with commutation relation $[X_1, X_2] = X_2$. The group is solvable,
since its Lie algebra has the ideal $\R X_2 = [\aff, \aff]$ with
$\aff/\R X_2$ abelian. Recall that a group~$G$ and its tangent Lie
algebra~$\g$ are called \textit{exponential} if the map
$\exp\: \g \to G$ is a surjective diffeomorphism. Note that
$$
\ad X = \twobytwo{0}{-v}{0}{u},  \word{with eigenvalues} 0,u.
$$
By an old result of Dixmier~\cite{DixmierExp}, since these eigenvalues
are not (nonzero) purely imaginary, the group is exponential. Of
course, this can be seen already from \eqref{eq:affine-group}, since
\begin{equation}
g(u,v) := \exp(uX_1 + vX_2) = \twobytwo{e^u}{v(e^u - 1)/u}{0}{1}.
\label{eq:affine-exp}
\end{equation}

Since $\tr(\ad X) \neq 0$ in general, the group is not unimodular.
Indeed, the right and left Haar measures $d_rg$ and $d_lg$ on $\Aff$
are respectively given by
\begin{subequations}
\label{eq:Haar}
\begin{align}
d_r(\exp X) = \det\biggl(\frac{e^{\ad X} - 1}{\ad X}\biggl) \,dX
= \frac{du\,dv}{\la(-u)} = \frac{da\,db}{a},
\label{eq:Haar-left}
\\[\jot]
d_l(\exp X) = \det\biggl(\frac{1 - e^{-\ad X}}{\ad X}\biggr) \,dX
= \frac{du\,dv}{\la(u)} = \frac{da\,db}{a^2},
\label{eq:Haar-right}
\end{align}
\end{subequations}
where $dX = du\,dv$ and
\begin{equation}
\la(t) := \frac{te^t}{e^t - 1} = \frac{t}{1 - e^{-t}}
= \frac{e^{t/2}}{\sinch(t/2)}
= e^{t/2}\,\Ga\bigl(1 + \tfrac{t}{2\pi i}\bigr)
\,\Ga\bigl(1 - \tfrac{t}{2\pi i}\bigr),
\label{eq:lambda-curve}
\end{equation}
with $\sinch t:= (\sinh t)/t$ (and $\sinch 0 := 1$), a well-known
nonvanishing even function (so named, by analogy with the
\textit{sinus cardinalis} of sampling theory). Note that the right
Haar measure on~$\Aff$ is the product of the Haar measures
of~$\R_+^\x$ and~$\R$; this is a general property for semidirect
products. Neither the left nor right Haar measure coincides with the
measure induced on~$\Aff$ by the Lebesgue measure on~$\aff$. The last
equalities on the right hand sides of~\eqref{eq:Haar} follow
from~\eqref{eq:affine-exp}. Therefore the \textit{modular function}
$\Dl(g) := d_l g/d_r g$ is given by $1/a$. We also note for future use
the product of the densities (normalized at~$0$) of the Haar measures
with respect to the Lebesgue measure,
\begin{equation}
j_l(X)j_r(X) := \frac{d_l(\exp X)}{dX}\,\frac{d_r(\exp X)}{dX}
= \sinch^2(u/2).
\label{eq:dreadful}
\end{equation}

The adjoint action of~$\Aff$ on~$\aff$, and the contragredient
coadjoint action on~$\aff^*$, are respectively given by
$$
\Ad g(X) := gXg^{-1} = \twobytwo{u}{av - bu}{0}{0};  \qquad
\bilin{g\lt F}{X} := \bilin{F}{\Ad g^{-1}(X)},
$$
for $X \in \aff$, $F \in \aff^*$; we generally use the letter~$F$
for points in coalgebras~$\g^*$. Also, $\aff^*$ can be realized by
matrices
$$
F = (x,y) := \biggl\{\twobytwo{x}{0}{y}{0} : (x,y) \in \R^2 \biggr\}.
$$
Thus $\bilin{F}{X} = \tr(FX) = ux + vy$, and the coadjoint action is
given by
$$
(x,y) \mapsto g \lt (x,y) \equiv \Coad g\,(x,y)
= \biggl( x + \frac{by}{a}, \frac{y}{a} \biggr).
$$
The orbits of this action are two open half-planes, plus an axis of
fixed points. Indeed, if we choose any point $F = (x,0)$, then the
whole group leaves it invariant, whereas all the other points
of~$\aff^*$ are found in the orbits of $F = (0,+1)$ and of
$F = (0,-1)$. The isotropy group in the last two cases is trivial, and
both orbits, which we denote by $\Oh_\pm$, are diffeomorphic to the
group itself. We think of $\Oh_\pm$ as Poincar\'e half-planes,
respectively $\Pi$ and $-\Pi$, adopting complex-variable notation when
convenient.

We can describe these ``solvmanifolds'' by group parameters. If
\begin{equation}
z(g) := g \lt \pm i = (\pm b/a, \pm 1/a),
\word{with inverse}  z \mapsto g_z = (\pm 1/y, x/y),
\label{eq:root-of-evil}
\end{equation}
then of course $g \lt z(g') = z(gg')$, for $g,g' \in \Aff$. One can
employ this to transfer the group operation onto the orbit by
$z(g) \. z(g') := z(gg')$. Explicitly,
\begin{equation}
(x + iy)\.(x' + iy') = x' \pm xy' \pm iyy'.
\label{eq:orbit-prod}
\end{equation}
Reciprocally, $z \mapsto g_z$ is an isomorphism, namely
$g_z\,g_{z'} = g_{z\.z'}$.

The invariant symplectic forms $\om_\pm$ on $\Oh_\pm$ are exact in the
present case, being clearly given by $\om_\pm(z) = dx\,dy/|y|$; note
that $\om_\pm(z(g)) = d_l g$. Darboux coordinates are $q = x/y$,
$p = |y|$.

In general, we say that a subalgebra $\h \subseteq \g$ is
\textit{subordinate} to $F \in \g^*$ if $F\bigr|_{[\h,\h]} = 0$ and
the map $X \mapsto \bilin{F}{X}$ is a one-dimensional representation
of~$\h$. The entire Lie algebra $\aff$ is subordinate to
each~$(x,0)$. Any one-dimensional subalgebra of~$\aff$ is subordinate
to $(0,1)$ or to $(0,-1)$, but only the ideal $[\aff, \aff]$ is
\textit{Puk\'anszky}, which means that $F + \h^\perp \subseteq \Oh_F$:
indeed, $(x,0) + (0,0) = (x,0)$ and $(0,\pm 1) + (x',0) \in \Oh_\pm$.

\subsection{The Kirillov map and the unirreps}
\label{ssc:unirreps}

The Kirillov theory asserts the existence of a map
$K\: \aff^*/\!\Aff \to \widehat\Aff$, where $\Aff$ acts via~$\Coad$,
the space $\aff^*/\!\Aff$ is endowed with the (non-Hausdorff, not even
$T_1$) quotient topology and the unitary dual $\widehat\Aff$ with its
standard Fell topology, determined by the hull-kernel topology on the
set of primitive ideals of $C^*(G)$ ---this matter is well explained
in \cite[Chap.~3]{DixmierCalg} or \cite[Chap.~7]{FellD}. For
exponential groups, $K$ has been proved by Leptin and
Ludwig~\cite{LeptinL} to be bijective and bicontinuous. (For the
similar correspondence between $\g^*/G$ and the set of primitive
ideals of the enveloping algebra $\U(\g)$, we refer
to~\cite{Olivier}.) It is known that all unirreps for exponential
groups are monomial, that is, induced by an abelian character of some
closed subgroup~$H$. If $H$ is the closed subgroup generated by $\h$
subordinate to~$F$, the Puk\'anszky condition guarantees that the
induced representations
$$
K[\Oh_F](\exp X) := \Ind_H^{\Aff}U_{F,H}(\exp X)
= \Ind_H^{\Aff}e^{2\pi i\bilin{F}{X}}
$$
are indeed irreducible. In the present case,
$$
U_{(x,0),\Aff}(\exp X) = e^{2\pi ixu}  \word{and}
U_{\pm,\exp(\R X_2)}(\exp bX_2) = e^{\pm 2\pi ib}.
$$
So we obtain in the first place the unitary one-dimensional
representations $(a,b) \mapsto a^{2\pi ix}$ of~$\Aff$; observe that
$K[\Oh_{(0,0)}]$ is the trivial representation.

Now denote $U_\pm := K[\Oh_\pm]$. The Kirillov scheme ``predicts''
(this is only of heuristic value) that the ``functional dimension''
of~$U_\pm$ is $\half \dim\Oh_\pm = 1$; that is, $U_\pm$ can be
realized on spaces of functions of one variable, in such a way that
the smooth vectors are smooth functions and the enveloping
algebra $U(\aff)$ acts by differential operators. The actual induction
process leads us to consider the space of functions $f$ on~$\Aff$ such
that
$$
f(a,b) = e^{\pm2\pi ib} \psi(a),
$$
and then, necessarily,
$$
U_\pm(a,b)f(a',b') = f(aa', ba' + b'),
\word{or} U_\pm(a,b)\psi(a') = e^{\pm2\pi iba'} \psi(aa').
$$
That is, we may settle on
$$
U_\pm(a,b) \psi(r) = e^{2\pi ibr} \psi(ar)
= U_\pm(1,b) U_\pm(a,0) \psi(r),
$$
with $r > 0$ for $U_+$ and $r < 0$ for $U_-$. The $U_\pm$ preserve the
space of smooth functions on the semiaxis, vanishing in some
neighbourhood of $r = 0$, and are unitary on the Hilbert spaces
$\K_\pm := L^2((0,\pm\infty), r^{-1}\,dr)$. Observe that $U_+$ and
$U_-$ are mutually dual; this is a general property for $K[\Oh]$ and
$K[-\Oh]$. (Needless to say, the process of going from $\Oh$ to a
putative $K[\Oh]$ is not always so smooth; experience points to the
importance of~$\Oh$ being spin and of its twisted Dirac operator to
construct $K[\Oh]$ ---see \cite{ConnesMHomog} in this respect.)

The selfadjoint infinitesimal generators for the unirreps are given by
\begin{align*}
U_\pm(X_1)\psi(r)
&= -i\frac{d}{du}\Bigr|_{u=0}\, U_\pm(\exp(u,0)) \psi(r)
= -ir\,\psi'(r);
\\
U_\pm(X_2)\psi(r)
&= -i\frac{d}{dv}\Bigr|_{v=0}\, U_\pm(\exp(0,v)) \psi(r)
= 2\pi r\,\psi(r).
\end{align*}
We denote the generators $U_\pm(X_1),\,U_\pm(X_2)$ by
$2\pi\hat\bt_\pm$, $2\pi\hat f_\pm$ respectively. Note
$2\pi i [\hat\bt_\pm,\hat f_\pm] = \hat f_\pm$.

Interesting (improper) denizens of $\K_\pm$ are the ``plane waves'' or
eigenfunctions of~$\hat\bt_\pm$. They are given by
$\psi_\bt(r) := r^{\pm2\pi i\bt}$, for $\bt$ real, and constitute a
(generalized) orthonormal and complete set. Complete orthonormal sets
within $\K_\pm$ are also known, but here we shall not use them.

\subsection{Characters of the unirreps and Kirillov's Fourier transform}
\label{ssc:FK}

Any operator~$A$ on $\K_\pm$ determines an integral kernel, with
suitable genuflections to rigour:
$$
A\psi(r) =: \int_{\R_\pm} A(r,s) \psi(s) \,\frac{ds}{s},
$$
so that the operator kernel of $U_\pm(a,b)$ is
$$
U_\pm(a,b;r,s) = e^{2\pi ibr} s\,\dl(s - ar).
$$

Let $f \in \D(\Aff) \equiv \Coo_c(\Aff)$. Following Kirillov
\cite[Sect.~4.1]{Kirillov}, we define the associated operators
$U_\pm(f)$ on~$\K_\pm$ by
$$
U_\pm(f)\psi(r) := \int_0^\infty \int_{-\infty}^\infty f(a,b)
\,e^{2\pi ibr} \psi(ar) \,\frac{da\,db}{a}
= \int_0^{\pm\infty} \int_{-\infty}^\infty f(s/r,b) \,e^{2\pi ibr}
\psi(s) \,\frac{dsdb}{s},
$$
with kernels
$$
U_\pm(f;r,s) = \int_{-\infty}^\infty f(s/r,b) \,e^{2\pi ibr} \,db,
$$
where $r > 0$, $s > 0$ or $r < 0$, $s < 0$, respectively. One would
na\"{\i}vely expect that $U_\pm(f)$ be nuclear, for $f$ a test
function. This is not the case unless $f$ is of zero mean with respect
to the second variable; otherwise $U_\pm(f)$ is not even
compact~\cite{Khalil}: $C^*(\Aff)$ is not \textit{liminaire}. Assume,
however, that this requirement holds; then the traces are given by
$$
\Tr U_\pm(f) = \int_{-\infty}^\infty \int_{\R_\pm^\x}
f(1,b) \,e^{2\pi ibr} \,\frac{db\,dr}{r}
$$
for $r > 0$ or $r < 0$, respectively. By definition, the (generalized)
characters $\chi_\pm$ of $U_\pm$ are the functionals such that
$$
\chi_\pm(f) = \Tr U_\pm(f).
$$
Let $\tilde f := f \circ \exp$. We see that $\chi_\pm(f)$ only depends
on the values of~$\tilde f$ on $[\aff,\aff]$; this is a particular
instance of a property established by Duflo~\cite{DufloCar}. We can
say that
\begin{equation}
\chi_\pm(a,b) = \dl(a - 1)\int_{\R_\pm^\x} e^{2\pi ibr}\,\frac{dr}{r},
\label{eq:Aff-chars}
\end{equation}
or $F_2\chi_\pm(a,r) = \frac{\th(\pm r)}{r}\,\dl(a - 1)$, where $\th$
is the Heaviside function, with the obvious caveats for good
definition in these ``ultraviolet divergent'' expressions.

For simply connected nilpotent groups, which are unimodular and whose
Haar measure is the Lebesgue measure in exponential coordinates,
Kirillov postulated and showed the existence of a unitary
transformation matching $L^2(G)$ with $L^2(\g^*)$, here denoted~$\FK$,
of the form
$$
\FK[f](F) := \int_\g \tilde f(X) \,e^{2\pi i\bilin{F}{X}} \,dX,
$$
and he established the formula
\begin{equation}
\chi_{K[\Oh]}(\exp X) = \int_\Oh e^{2\pi i\bilin{F}{X} + \om},
\label{eq:Kiri-dixit}
\end{equation}
where $\om$ is the invariant symplectic form on~$\Oh$, so that
\begin{equation}
\Tr K[\Oh](f) := \int_\g f(\exp X)\,\chi_{K[\Oh]}(\exp X) \,dX
= \int_\Oh \FK[f](F) \, d\mu_\om(F).
\label{eq:Russian-ideology}
\end{equation}
Here $\mu_\om$ is the Liouville measure on the orbit, given by
$\om^{\half\dim\Oh}/(\half\dim\Oh)!\,$; also,
\eqref{eq:Russian-ideology} clearly means that $\FK[\chi_{K[\Oh]}]$
---a tempered distribution defined on~$\g$ by transposition,
transported to~$G$ by the exponential map--- coincides precisely with
that measure. This was one of the earliest triumphs of the method of
orbits. The similarity of \eqref{eq:Kiri-dixit} with the relation
between the classical and quantum partition functions has been pointed
out and exploited before: see~\cite{Rossmann}.

For general solvable groups, this will not do. Part of the problem is
that the orbit is open and the invariant symplectic structure is
singular at the boundary. Also, it is unclear which measure to use
on~$\g$, in the nonunimodular case. Kirillov suggests that the recipe
\eqref{eq:Kiri-dixit} be replaced by a weighted version:
$$
\chi_{K[\Oh]}(\exp X)
= \frac{1}{q(X)}\int_\Oh e^{2\pi i\bilin{F}{X} + \om}
$$
and for $q$ he chooses $(j_lj_r)^{1/4}$. In view
of~\eqref{eq:dreadful}, for~$\Aff$ this leads to
\begin{align}
\FK[f](z) &= \iint_{\R^2} \exp\{2\pi i(ux + vy)\}
\, f\biggl(e^u, \frac{v}{\la(-u)} \biggr)
\,\frac{du\,dv}{\sqrt{\sinch\frac{u}{2}}} 
\nonumber \\
&= \iint_{\Aff} \exp\{2\pi i(x\log a + yb\la(-\log a))\}\, f(a,b)\,
\la^{1/4}(\log a) \la^{5/4}(-\log a) \,\frac{da\,db}{a},
\label{eq:quite-dreadful}
\end{align}
for $z \in \aff^*$. It is necessary and sufficient that $\FK[f]$ go to
zero on the boundary of the orbits for $K[\Oh_\pm](f)$ to be nuclear.
Then~\eqref{eq:Russian-ideology} is still valid. This is scarcely
surprising since, as pointed out earlier, \textit{only} the value
of~$\la$ at~$0$ enters the calculation. However, the last formula is
certainly not pretty.

Yet another \textit{a priori} Fourier map is defined in~\cite{AliFK}
by a formula of the type
\begin{equation}
\FAFK[f](F) = \sqrt{\mathstrut\smash{\xi_j(F)}} \int_{\g^*}
\exp\{2\pi i\bilin{F}{X}\}\, f(\exp X) \,\sqrt{m(X)} \,dX.
\label{eq:woe-to-come}
\end{equation}
We explain their notation: $j$ is an index parametrizing the orbits;
$\xi_j = dF/d\om_j$; and $m(X) = d_l(\exp X)/dX$. Here, for
$y \neq 0$, we take $j \in \{\pm\}$; $\xi_\pm(z) = |y|$; and
$m(u,v) = 1/\la(u)$. Hence
\begin{equation}
\FAFK[f](z) = \sqrt{|y|} \iint_{\R^2} \exp\{2\pi i(ux + vy)\}\,
f(e^u, v/\la(-u)) \,\frac{1}{\sqrt{\la(u)}} \,du\,dv.
\label{eq:Four-Fuhr}
\end{equation}
By construction, this is an isometry between $L^2(\Aff, d_lg)$ and
$L^2(\aff^*, d\om_+ \cup d\om_-)$. However, it does not give the
character.

In this paper we introduce two (left and right) powerful alternative
transforms to Kirillov's Fourier map, that likewise recover the
character, and are closely related to~$\FAFK$.

\subsection{The Duflo--Moore operators}
\label{ssc:stretch}

The decomposition of the regular representation of~$\Aff$, defined as 
usual by
$$
\La(g)f(g') := f(g^{-1}g'),
$$
is well known for the affine group: $\La$ decomposes into a continuous
direct sum of representations equivalent to $U_+ \oplus U_-$. More
concretely, the Plancherel measure is in our case just the counting
measure on the two-element set $\{U_\pm\}$, and there is a unitary map
$P$, the \textit{Plancherel transform}:
$$
P : L^2(G) \to \HS(\K_+) \oplus \HS(\K_-),
$$
with $\HS(\K_\pm)$ denoting the Hilbert algebras of Hilbert--Schmidt
operators on~$\K_\pm$, and $P$ given~by
\begin{equation}
Pf := U_+(f)\,d_+^{1/2} \oplus U_-(f)\,d_-^{1/2},
\label{eq:Planch-punch}
\end{equation}
where the $d_\pm$ are positive operators on~$\K_\pm$ with densely
defined inverses, determined (up to a positive constant) by the
semi-invariance relation:
\begin{equation}
U_\pm(g)\, d_\pm \,U^\7_\pm(g) = \Dl^{-1}(g) \,d_\pm.
\label{eq:semi-invce}
\end{equation}
For a general proof of this uniqueness for $G$ not unimodular, see
\cite{DufloM} and also~\cite{Aniello}. Because the $U_\pm$ are induced
from the subgroup $\{1,b\}$ belonging to the kernel of~$\Dl$, it is
easily seen that
$$
d_\pm \psi(r) = |r|\,\psi(r),
$$
where we have chosen a convenient normalization.  These $d_\pm$ are
the formal dimension operators as originally defined by Duflo and
Moore in~\cite{DufloM}, although later authors use the phrase
``Duflo--Moore operators'' for $d_\pm^{-1/2}$ instead.  For their
theory, one may consult~\cite{DufloM} and also its excellent
precursor~\cite{Tatsuuma}.  The remarkable thing is that the operators
$U_\pm(f)\,d^{1/2}_\pm$ (or rather, their closures) are
Hilbert--Schmidt whenever $f$ belongs to $L^2(G)$ ---actually, our
treatment of the harmonic analysis on~$\Aff$ in the long
Section~\ref{sec:FM-Aff} amounts to an indirect proof of this fact.
Then \eqref{eq:Planch-punch} holds, and moreover
$$
P(\La(g) f) = \bigl( U_+ \oplus U_- \bigr)(g)\, Pf.
$$
Also, for $f$ in a suitable dense subspace of $L^2(G)$, the operators
$U(f)\,d_\pm$ are nuclear.

\section{An unusual special function}
\label{ssc:lambda}

Before we plunge into calculating the quantizer, it is convenient to
perform a few exercises in real analysis, to be rewarded with later
simplification.

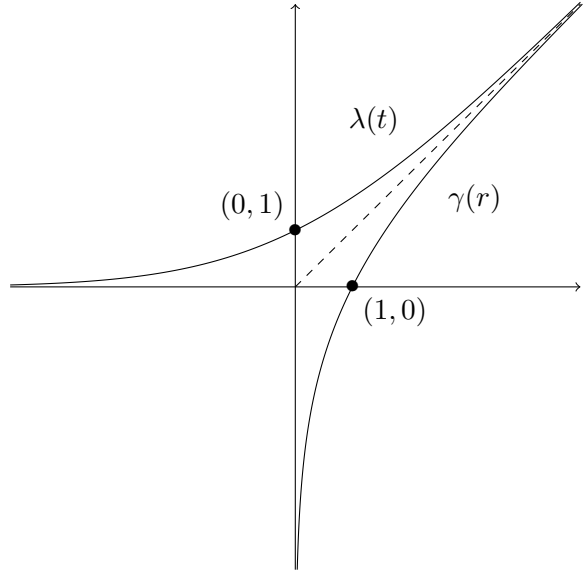
\begin{figure}[htb]
\centering
\begin{tikzpicture}[scale=0.75]
\draw[dashed] (0,0) -- (5,5);
\draw[->] (-5,0) -- (5,0);
\draw[->] (0,-5) -- (0,5);
\draw (0,1) node {$\8$} node[above left] {$(0,1)$};
\draw (2,2.5) node[above left] {$\la(t)$};
\draw plot coordinates{ 
(-5,0.033918)
(-4.9,0.036762)
(-4.8,0.039831)
(-4.7,0.04314)
(-4.6,0.046708)
(-4.5,0.050552)
(-4.4,0.054692)
(-4.3,0.059147)
(-4.2,0.06394)
(-4.1,0.069093)
(-4,0.074629)
(-3.9,0.080574)
(-3.8,0.086954)
(-3.7,0.093796)
(-3.6,0.10113)
(-3.5,0.10898)
(-3.4,0.11739)
(-3.3,0.12638)
(-3.2,0.13598)
(-3.1,0.14624)
(-3,0.15719)
(-2.9,0.16886)
(-2.8,0.18129)
(-2.7,0.19453)
(-2.6,0.20861)
(-2.5,0.22356)
(-2.4,0.23945)
(-2.3,0.25629)
(-2.2,0.27414)
(-2.1,0.29304)
(-2,0.31304)
(-1.9,0.33416)
(-1.8,0.35646)
(-1.7,0.37998)
(-1.6,0.40475)
(-1.5,0.43083)
(-1.4,0.45824)
(-1.3,0.48702)
(-1.2,0.51722)
(-1.1,0.54886)
(-1,0.58198)
(-0.9,0.61661)
(-0.8,0.65277)
(-0.7,0.6905)
(-0.6,0.72982)
(-0.5,0.77075)
(-0.4,0.8133)
(-0.3,0.85749)
(-0.2,0.90333)
(-0.1,0.95083)
(0,1)
(0.1,1.0508)
(0.2,1.1033)
(0.3,1.1575)
(0.4,1.2133)
(0.5,1.2707)
(0.6,1.3298)
(0.7,1.3905)
(0.8,1.4528)
(0.9,1.5166)
(1,1.582)
(1.1,1.6489)
(1.2,1.7172)
(1.3,1.787)
(1.4,1.8582)
(1.5,1.9308)
(1.6,2.0048)
(1.7,2.08)
(1.8,2.1565)
(1.9,2.2342)
(2.0,2.313)
(2.1,2.393)
(2.2,2.4741)
(2.3,2.5563)
(2.4,2.6394)
(2.5,2.7236)
(2.6,2.8086)
(2.7,2.8945)
(2.8,2.9813)
(2.9,3.0689)
(3,3.1572)
(3.1,3.2462)
(3.2,3.336)
(3.3,3.4264)
(3.4,3.5174)
(3.5,3.609)
(3.6,3.7011)
(3.7,3.7938)
(3.8,3.887)
(3.9,3.9806)
(4,4.0746)
(4.1,4.1691)
(4.2,4.2639)
(4.3,4.3591)
(4.4,4.4547)
(4.5,4.5506)
(4.6,4.6467)
(4.7,4.7431)
(4.8,4.8398)
(4.9,4.9368)
(5.0,5.0339) };
\draw (1,0) node {$\8$} node[below right] {$(1,0)$};
\draw (2.5,2) node[below right] {$\ga(r)$};
\draw plot coordinates{ 
(0.033918,-5)
(0.036762,-4.9)
(0.039831,-4.8)
(0.04314,-4.7)
(0.046708,-4.6)
(0.050552,-4.5)
(0.054692,-4.4)
(0.059147,-4.3)
(0.06394,-4.2)
(0.069093,-4.1)
(0.074629,-4)
(0.080574,-3.9)
(0.086954,-3.8)
(0.093796,-3.7)
(0.10113,-3.6)
(0.10898,-3.5)
(0.11739,-3.4)
(0.12638,-3.3)
(0.13598,-3.2)
(0.14624,-3.1)
(0.15719,-3)
(0.16886,-2.9)
(0.18129,-2.8)
(0.19453,-2.7)
(0.20861,-2.6)
(0.22356,-2.5)
(0.23945,-2.4)
(0.25629,-2.3)
(0.27414,-2.2)
(0.29304,-2.1)
(0.31304,-2)
(0.33416,-1.9)
(0.35646,-1.8)
(0.37998,-1.7)
(0.40475,-1.6)
(0.43083,-1.5)
(0.45824,-1.4)
(0.48702,-1.3)
(0.51722,-1.2)
(0.54886,-1.1)
(0.58198,-1)
(0.61661,-0.9)
(0.65277,-0.8)
(0.6905,-0.7)
(0.72982,-0.6)
(0.77075,-0.5)
(0.8133,-0.4)
(0.85749,-0.3)
(0.90333,-0.2)
(0.95083,-0.1)
(1,0)
(1.0508,0.1)
(1.1033,0.2)
(1.1575,0.3)
(1.2133,0.4)
(1.2707,0.5)
(1.3298,0.6)
(1.3905,0.7)
(1.4528,0.8)
(1.5166,0.9)
(1.582,1)
(1.6489,1.1)
(1.7172,1.2)
(1.787,1.3)
(1.8582,1.4)
(1.9308,1.5)
(2.0048,1.6)
(2.08,1.7)
(2.1565,1.8)
(2.2342,1.9)
(2.313,2)
(2.393,2.1)
(2.4741,2.2)
(2.5563,2.3)
(2.6394,2.4)
(2.7236,2.5)
(2.8086,2.6)
(2.8945,2.7)
(2.9813,2.8)
(3.0689,2.9)
(3.1572,3)
(3.2462,3.1)
(3.336,3.2)
(3.4264,3.3)
(3.5174,3.4)
(3.609,3.5)
(3.7011,3.6)
(3.7938,3.7)
(3.887,3.8)
(3.9806,3.9)
(4.0746,4)
(4.1691,4.1)
(4.2639,4.2)
(4.3591,4.3)
(4.4547,4.4)
(4.5506,4.5)
(4.6467,4.6)
(4.7431,4.7)
(4.8398,4.8)
(4.9368,4.9)
(5.0339,5) } ;
\end{tikzpicture}
\caption{The function $\la$ and its inverse function $\ga$}
\label{fg:lambda}
\end{figure}

We begin with the function $\la$ of~\eqref{eq:lambda-curve}. Note that
$\la(0) = 1$ and $\la(t) > 0$ for all $t \in \R$, that
$\la(t) \downto 0$ as $t \to -\infty$, and that $\la(t) \sim t$ as
$t \to +\infty$ (see Figure~\ref{fg:lambda}). It is easy to see that
\begin{subequations}
\label{eq:la-relns}
\begin{align}
\la(-t) &= e^{-t} \la(t),
\label{eq:la-reln-mult}
\\
\la(t) - \la(-t) &= t.
\label{eq:la-reln-add}
\end{align}
\end{subequations}
These functional equations determine $\la$ uniquely. It is an
analytic function for $|t| < 2\pi$, with expansion
$$
\la(t) = \sum_{n\geq 0} (-1)^n B_n \frac{t^n}{n!};
\word{thus} \la(-t) = \sum_{n\geq0} B_n \frac{t^n}{n!},
$$
where the $B_n$ are the well-known Bernoulli numbers.

The derivative of $\la(t)$ is
\begin{align}
\la'(t) &= (1 - e^{-t})^{-1} - te^{-t} (1 - e^{-t})^{-2}
= \frac{\la(t)}{t} - \frac{\la(t)\la(-t)}{t}
\nonumber \\
&= \frac{\la(t)}{t}\, (1 - \la(-t))
= \frac{\la(t)}{t}\, (1 + t - \la(t)).
\label{eq:la-deriv}
\end{align}

Thus $\la(t)$ is strictly increasing on~$\R$. Since
\eqref{eq:la-reln-add} entails $\la'(t) + \la'(-t) = 1$, we see that
$\la'(0) = \half$. A brief calculation shows that $\la''(t) > 0$ for
$t > 0$; since $\la''(t) = \la''(-t)$, it follows that $\la'(t)$ is
increasing (i.e., $\la(t)$ is convex), with $\la'(t) \uparrow 1$ as
$t \to +\infty$, and therefore $\la'(t) < 1$ for all~$t$.

Next, let $\ga(r)$ denote the inverse function for~$\la$: $\ga(r) = t$
when $r = \la(t)$. It is defined for~$r > 0$, is strictly increasing
with $\ga(1) = 0$ and $\ga'(1) = 2$, and $\ga(r)\downto-\infty$ as
$r\downto0$, while $\ga(r) \sim r$ as $r \uparrow +\infty$. All that
is evident on reflecting the graph of $r = \la(t)$ about the main
diagonal $r = t$. Of course, the chain rule and \eqref{eq:la-deriv}
give
$$
\ga'(r) = \frac{1}{\la'(\ga(r))} = \frac{\ga(r)}{r(1 - r + \ga(r))}.
$$
In particular, $\ga'(r) > 1$ for all $r > 0$.

The special function worthy of our attention is
$$
\sg(r) := r - \ga(r),  \word{for}  r > 0.
$$
For $|r| < 1$ we obtain, by \cite[Sect.~2.1]{Quaoar} for instance,
\begin{align*}
\ga(1 + r)
&= 2r - \frac{2}{3} r^2 + \frac{4}{9} r^3 - \frac{44}{135} r^4
+ \frac{104}{405} r^5 +\cdots
\\[\jot]
\word{and} \sg(1 + r)
&= 1 - r + \frac{2}{3} r^2 - \frac{4}{9} r^3 + \frac{44}{135} r^4
- \frac{104}{405} r^5 +\cdots
\end{align*}

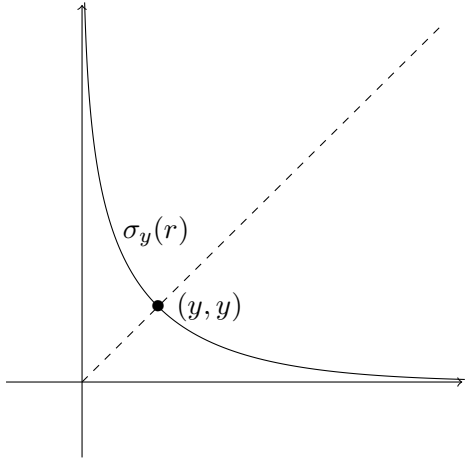
\begin{figure}[htb]
\centering
\begin{tikzpicture}
\draw[dashed] (0,0) -- (4.7,4.7);
\draw[->] (-1,0) -- (5,0);
\draw[->] (0,-1) -- (0,5);
\draw (1,1) node {$\8$} node[right=3pt] {$(y,y)$};
\draw (0.4,2) node[right] {$\sg_y(r)$};
\draw plot coordinates{ 
(0.033918,5.0339)
(0.036762,4.9368)
(0.039831,4.8398)
(0.04314,4.7431)
(0.046708,4.6467)
(0.050552,4.5506)
(0.054692,4.4547)
(0.059147,4.3591)
(0.06394,4.2639)
(0.069093,4.1691)
(0.074629,4.0746)
(0.080574,3.9806)
(0.086954,3.887)
(0.093796,3.7938)
(0.10113,3.7011)
(0.10898,3.609)
(0.11739,3.5174)
(0.12638,3.4264)
(0.13598,3.336)
(0.14624,3.2462)
(0.15719,3.1572)
(0.16886,3.0689)
(0.18129,2.9813)
(0.19453,2.8945)
(0.20861,2.8086)
(0.22356,2.7236)
(0.23945,2.6394)
(0.25629,2.5563)
(0.27414,2.4741)
(0.29304,2.393)
(0.31304,2.313)
(0.33416,2.2342)
(0.35646,2.1565)
(0.37998,2.08)
(0.40475,2.0048)
(0.43083,1.9308)
(0.45824,1.8582)
(0.48702,1.787)
(0.51722,1.7172)
(0.54886,1.6489)
(0.58198,1.582)
(0.61661,1.5166)
(0.65277,1.4528)
(0.6905,1.3905)
(0.72982,1.3298)
(0.77075,1.2707)
(0.8133,1.2133)
(0.85749,1.1575)
(0.90333,1.1033)
(0.95083,1.0508)
(1,1)
(1.0508,0.95083)
(1.1033,0.90333)
(1.1575,0.85749)
(1.2133,0.8133)
(1.2707,0.77075)
(1.3298,0.72982)
(1.3905,0.6905)
(1.4528,0.65277)
(1.5166,0.61661)
(1.582,0.58198)
(1.6489,0.54886)
(1.7172,0.51722)
(1.787,0.48702)
(1.8582,0.45824)
(1.9308,0.43083)
(2.0048,0.40475)
(2.08,0.37998)
(2.1565,0.35646)
(2.2342,0.33416)
(2.313,0.31304)
(2.393,0.29304)
(2.4741,0.27414)
(2.5563,0.25629)
(2.6394,0.23945)
(2.7236,0.22356)
(2.8086,0.20861)
(2.8945,0.19453)
(2.9813,0.18129)
(3.0689,0.16886)
(3.1572,0.15719)
(3.2462,0.14624)
(3.336,0.13598)
(3.4264,0.12638)
(3.5174,0.11739)
(3.609,0.10898)
(3.7011,0.10113)
(3.7938,0.093796)
(3.887,0.086954)
(3.9806,0.080574)
(4.0746,0.074629)
(4.1691,0.069093)
(4.2639,0.06394)
(4.3591,0.059147)
(4.4547,0.054692)
(4.5506,0.050552)
(4.6467,0.046708)
(4.7431,0.04314)
(4.8398,0.039831)
(4.9368,0.036762)
(5.0339,0.033918) } ;
\end{tikzpicture}
\caption{The self-inverse function $\sg_y$ for $y > 0$}
\label{fg:sigma}
\end{figure}

Write $s = \la(-t)$ and $r = \la(t)$. Then relation
\eqref{eq:la-reln-add} shows that
\begin{equation}
s = \sg(r)  \iff  r = \sg(s),
\label{eq:sg-recip}
\end{equation}
or, not to put too fine a point on it, $\sg(\sg(r)) = r$, that is,
$\sg$ is \textit{self-inverse}.

Note, too, that $\sg(r) = re^{-\ga(r)}$, in view of relation
\eqref{eq:la-reln-mult}. Also, $\sg(1) = 1$ and $\sg'(1) = -1$, as
expected since the graph of~$\sg$ must be invariant on reflection in
the main diagonal. Now $\sg'(r) = 1 - \ga'(r) < 0$ always, so that
$\sg$ is a strictly decreasing function. Moreover,
$\sg(r) \sim - \ga(r)$ as $r \downto 0$, while $\sg(r) \sim r e^{-r}$
as $r \to +\infty$: the graph of~$\sg$ is exponentially asymptotic to
both axes in the first quadrant. Note, as well,
from~\eqref{eq:sg-recip}:
$$
\ga(r) = - \ga(\sg(r)); \qquad \ga'(r) = - \ga'(\sg(r))\,\sg'(r).
$$
We remark that
$\sg'(r) = e^{-\ga(r)}(1 - r\ga'(r)) = \sg(r)(r^{-1} - \ga'(r))$;
writing $s = \sg(r)$ gives
\begin{equation}
s - \sg'(\sg(s))\,\sg(s)
= s - s\biggl( \frac{1}{\sg(s)} - \ga'(\sg(s)) \biggr) \sg(s)
= s \sg(s) \ga'(\sg(s)).
\label{eq:sg-trick}
\end{equation}

It is also useful for our purposes to dilate $\ga$ and $\sg$ by a
factor $y > 0$, as follows:
$$
\ga_y(r) := y\,\ga(r/y),  \qquad  \sg_y(r) := y\,\sg(r/y).
$$
It is immediate that $\sg_y(\sg_y(r)) = r$, so that $\sg_y$ also has a
graph that is unchanged on reflection in the diagonal, except that it
now crosses the diagonal at $(y,y)$: see Figure~\ref{fg:sigma}.
Moreover, the previous relations between~$\ga$ and~$\sg$ become
\begin{align}
\sg_y(r) &= r - \ga_y(r), & \ga_y(\sg_y(r)) &= - \ga_y(r),
\nonumber \\
\sg'_y(r) &= 1 - \ga'_y(r), & \ga'_y(\sg_y(r)) \,\sg'_y(r)
&= - \ga'_y(r).
\label{eq:gay-sgy-relns}
\end{align}

\section{The Stratonovich--Weyl quantizer for the `$ax + b$' group}
\label{sec:quantizer}

\subsection{Weyl quantization for the half-plane}
\label{ssc:Weyl}

Pre-existing work on image processing
\cite{BertrandsAff,BertrandsSymb} points the way towards the
quantizer: the transformation from ``density matrices'' to ``Wigner
functions'' considered in these papers respects reality, covariance
and ``unitarity'', which is equivalent to traciality. Consider the
\textit{Weyl operator},
$$
W_\pm(u,v) := \exp\{2\pi i(u{\hat\bt}_\pm + v{\hat f}_\pm)\} =
U_\pm(e^u,v(e^u - 1)/u) = U_\pm(e^u, ve^u/\la(u)).
$$
Our candidate quantizer is
\begin{equation}
\Om_\pm(x + iy)
:= |y|\iint_{\R^2} \exp\{-2\pi i(ux + vy)\}\, W_\pm(u,v) \,du\,dv.
\label{eq:Weyl-Ansatz}
\end{equation}
It can hardly be simpler! Aside from the factor $|y|$, that
compensates for the measures used on~$\Oh_\pm$, it means that the
familiar definition from Quantum Mechanics works. The recipe is
actually imposed on us by the heuristic rule that $\Om_\pm(F_0)$
should be the quantization of $\dl(F-F_0)$, or the equivalent remark
in~\cite{BertrandsSymb} that the quantization of a plane wave should
be given by the Weyl operator. It stands to reason that the Weyl
operator will play an essential role in an exponential group; one may
treat \eqref{eq:Weyl-Ansatz} as an \textit{Ansatz}, and simply prove
that it gives the Stratonovich--Weyl quantizer by verifying all the
required properties.

We carry out this verification for the upper half-plane~$\Pi$.
For~$-\Pi$ the argument is identical. It is no longer worth the
trouble to keep always the subscripts~$\pm$, so we mostly drop them.

\subsection{Identification of the basic ``parity'' operators}
\label{ssc:parity}

The claim is that we can associate to each symbol $f(z)$ on~$\Pi$ an
operator $A$ on the representation space $\K_+ \equiv \K$, and vice
versa, by
\begin{equation}
A = \int_\Pi \Om(z)\, f(z) \,d\om(z) =: Q(f), \qquad
f(z) = \Tr\bigl( \Om(z)A \bigr),
\label{eq:vice-versa}
\end{equation}
with the properties required in the Introduction. The quantizer
$\Om(z)$ remains to be determined.

With the machinery assembled in the previous section, the computation
of~$\Om$ is straightforward. First of all,
$$
W(u,v)\psi(r) = e^{2\pi iv e^u r/\la(u)} \,\psi(e^u r),
$$
with operator kernel
$$
W(u,v;r,s) = e^{2\pi ivs/\la(u)} s\,\dl(s - e^u r).
$$
The right hand side of~\eqref{eq:Weyl-Ansatz}, applied to
$\psi \in \D(0,\infty)$, yields
\begin{align}
\Om(z)\psi(r)
&= y \iint_{\R^2} e^{-2\pi i(ux + vy - e^urv/\la(u))}\, \psi(e^ur)
\,du\,dv
\nonumber \\
&= y \int_\R e^{-2\pi iux} \,\dl(y - re^u/\la(u)) \,\psi(e^ur) \,du
\nonumber \\
&= \frac{r}{y} \,\ga'_y(r)\,e^{2\pi ix\ga_y(r)/y}\,\psi(\sg_y(r)).
\label{eq:omz-oper}
\end{align}
For this, consider the diffeomorphism $u \mapsto r/\la(-u) = w$, whose
inverse mapping is $u = -\ga(r/w)$ with Jacobian
$J(u;w) = rw^{-2}\ga'(r/w)$. The general formula
\cite[Chap.~1]{VladimirovDZ}:
$$
\bigl<T(w(u)), \psi(u)\bigr> = \bigl<T(w), \psi(u(w))\,|J(u;w)|\bigr>
$$
in the present case easily gives
$$
\int_\R \dl(y - re^u/\la(u)) \, e^{-2\pi iux} \, \psi(e^ur) \, du
= \int_\R \dl(y - w) \, e^{2\pi ix\ga(r/w)} \, \psi(re^{-\ga(r/w)})
\,rw^{-2}\ga'(r/w) \,dw,
$$
yielding \eqref{eq:omz-oper}. Similar computations will reappear
throughout this paper; often we shall just omit them. In particular,
we have obtained
$$
\Om(i)\psi(r) = r \ga'(r) \,\psi(\sg(r)).
$$
A remarkable event has occurred. For the ordinary Moyal family, the
``mother'' operator in the Schr\"odinger representation is known to be
essentially the (Grossmann--Royer) parity operator. Matters are more
involved here, but still $\Om(i)$ is basically given by a kind of
reflection; to wit, the involutive $\sg$~function.

In view of~\eqref{eq:orbit-prod}, covariance requires
$\Om(z) = U(g_z) \Om(i) U^\7(g_z)$, where $U(g_z) = U(1/y,x/y)$ with
adjoint operator $U^\7(g_z) = U(y,-x)$. We leave this direct
verification as an exercise.

The kernel of~$\Om(z)$ is given by
\begin{subequations}
\label{eq:omz-ker}
\begin{equation}
\Om(z;r,s)
= \frac{rs}{y}\, \ga'_y(r) \,e^{2\pi ix(r - s)/y} \,\dl(s - \sg_y(r)).
\label{eq:omz-ker-joe}
\end{equation}
An alternative form is
\begin{equation}
\Om(z;r,s) = y\,e^{2\pi ix\log(r/s)}
\,\dl\biggl(y - \frac{r}{\la(\log(r/s))} \biggr)
= y\,e^{2\pi ix\log(r/s)}
\,\dl\biggl(y - \frac{r - s}{\log(r/s)} \biggr).
\label{eq:omz-ker-vic}
\end{equation}
\end{subequations}

For the normalization property the operators $\Om(z)$ must be of
trace~$1$, in the distributional sense. We check this by computing
\begin{equation}
\Tr\Om(i) = \int_0^\infty \Om(i;r,r) \,\frac{dr}{r}
= \int_0^\infty r \ga'(r) \,\dl(\ga(r)) \,dr = 1.
\label{eq:omi-trace}
\end{equation}
The normalization $\Tr\Om(z) = 1$ is then automatic, since the
$U(g_z)$ are unitary. It can be proved that the operators $Q(f)$, for
$f$ any test function on~$\Pi$, are nuclear~\cite{Eris}.

\subsection{Selfadjointness of the quantizer operators}
\label{ssc:selfadj}

Note that the $\Om(z)$ are unbounded operators. They are defined at
least on $\D(0,\infty)$. Indeed, since $\sg_y$ is monotonic and
smooth, the right hand side of~\eqref{eq:omz-oper} lies in
$\D(0,\infty)$ whenever $\psi$ does, and we see that
\begin{align*}
\|\Om(z)\psi\|^2
&= \frac{1}{y^2} \int_0^\infty r\ga'_y(r)^2\,|\psi(\sg_y(r))|^2 \,dr
= \frac{1}{y^2} \int_0^\infty \sg_y(s)
\ga'_y(\sg_y(s))^2 \,|\psi(s)|^2 \,|\sg'_y(s)| \,ds
\\
&= \frac{1}{y^2} \int_0^\infty s \sg_y(s) \ga'_y(s) \ga'_y(\sg_y(s))
\,|\psi(s)|^2\,\frac{ds}{s},
\end{align*}
where \eqref{eq:gay-sgy-relns} has been used. Since
$s \mapsto s\ga'_y(s)$ increases from $1$ to~$+\infty$ for
$0 < s < \infty$, no bound of the form $\|\Om(i)\psi\| \leq C\|\psi\|$
is possible. We remark that the equivalent operators are (of course)
bounded for compact groups, and for the Heisenberg group. But
unboundedness is not unheard~of, as it happens for the Poincar\'e
group \cite[Sect.~4]{Ganymede}.

The $\Om(z)$ are hermitian on the domain $\D(0,\infty)$. For
$\phi,\psi$ in this domain, we get
\begin{align}
\braket{\phi}{\Om(i)\psi}
&= \int_0^\infty \owl{\phi(r)}\, \Om(i)\,\psi(r) \,\frac{dr}{r}
= \int_0^\infty \owl{\phi(r)}\, \ga'(r) \,\psi(\sg(r)) \,dr
\label{eq:hermiticity} \\
&= \int_0^\infty \owl{\phi(\sg(s))}\, \ga'(\sg(s)) \,\psi(s)
\,|\sg'(s)|\,ds = \int_0^\infty s \ga'(s) \,\owl{\phi(\sg(s))}\,
\psi(s) \,\frac{ds}{s} = \braket{\Om(i)\phi}{\psi}.
\nonumber
\end{align}
Likewise, $\braket{\phi}{\Om(z)\psi} = \braket{\Om(z)\phi}{\psi}$ by
covariance, since $U^\7(g_z)$ preserves $\D(0,\infty)$. We prove that
$\Om(z)$ is closable and identify its closure, and then show
selfadjointness. This is easy to do using the fortunate fact that its
square is a multiplication operator, $\Om^2(z) = M_{\eta^2_z}$, where
the unbounded positive function $\eta_z$ is given by
$$
\eta^2_z(r) := r y^{-2} \sg_y(r)\,\ga'_y(r)\,\ga'_y(\sg_y(r)).
$$
Notice that $\eta_z(r) = \eta_i(r/y)$ for $z \in \Pi$. The natural
domains for the several $\Om(z)$ are the dense subspaces $B_z$ of $\K$
defined as
$$
B_z := \set{\psi \in \K : \eta_z \psi \in \K}.
$$
Note that \eqref{eq:hermiticity} remains valid for $\phi,\psi \in B_z$
so that $\Om(z)$ is hermitian on this domain.

\begin{prop}
On the respective domains $B_z$, the operators $\Om(z)$ are
selfadjoint.
\end{prop}

\begin{proof}
First, take $\psi \in \D(0,\infty)$. One sees at once that indeed
$\Om^2(z)\psi(r) = \eta^2_z(r)\,\psi(r)$. We already showed that
$\|\Om(z)\psi\| = \|\eta_z\psi\|$, using~\eqref{eq:omz-oper}; and this
continues to hold for all $\psi \in B_z$. Observe that $B_z$ is
complete in the graph norm given by
$\snorm{\psi}^2 := \|\psi\|^2 + \|\eta_z\psi\|^2$, and that
$\D(0,\infty)$ is dense in $B_z$ for this norm. Thus $\Om(z)$, with
domain $B_z$, is a closed operator. Clearly
$\Dom \Om(z) \subset \Dom \Om(z)^\7$ since $\Om(z)$ is hermitian
on~$B_z$.

Note that if $\psi \in \Dom \Om^\7(z)$, and if $\chi_n$ is the
indicator function of the interval $[1/n,n]$, say, then
$\chi_n \Om(z)^\7 \psi \in B_z$ and a routine argument, using the
monotone convergence theorem, shows that
$$
\|\Om(z)^\7 \psi\| = \lim_{n\to\infty} \|\chi_n \Om(z)^\7 \psi\|
= \lim_{n\to\infty} \sup_{\|\phi\|=1}
|\braket{\Om(z)(\chi_n\phi)}{\psi}|
= \cdots = \|\eta_z\psi\|,
$$
so that necessarily $\psi \in B_z$. Thus
$\Dom \Om(z)^\7 = B_z = \Dom \Om(z)$, as required.

Note also that the original domain $\D(0,\infty)$, being dense in each
$B_z$ for the graph norm, is a common core for all the $\Om(z)$, which
are therefore essentially selfadjoint on that domain.
\end{proof}

A consequence of the operators $\Om(z)$ being selfadjoint (not just
formally so) is that they become \textit{observables} in the
quantum-mechanical sense. For the Stratonovich--Weyl operators in the
standard (Heisenberg covariant) case, see the discussion
in~\cite{Royer}.

\subsection{Traciality}
\label{ssc:tracial}

\begin{lem}
\label{lm:tracial}
The quantizer is \emph{tracial}, in the sense that
$$
\Tr\bigl( \Om(w)\Om(z) \bigr) = y\,\dl(w - z)
\word{for all} w,z \in \Pi,
$$
where the right hand side is the reproducing kernel for the Hilbert
space $L^2(\Pi, dx\,dy/y)$.
\end{lem}

\begin{proof}
With $w = u + iv$, $z = x + iy$, we obtain
\begin{align*}
\Tr\bigl(\Om(w)\Om(z)\bigr)
&= \int_0^\infty \int_0^\infty \Om(w;r,s)\Om(z;s,r)\,\frac{dr\,ds}{rs}
\\
&= \int_0^\infty \int_0^\infty \frac{rs}{vy}\,
\ga'\Bigl(\frac{s}{v}\Bigr) \,\ga'\Bigl(\frac{r}{y}\Bigr) \, e^{-2\pi
i(r - s)(x/y - u/v)} \,\dl(r - \sg_v(s)) \,\dl(s - \sg_y(r))\,dr\,ds
\\
&= \int_0^\infty \frac{r}{v} \sg\Bigl(\frac{r}{y}\Bigr)
\,\ga'\biggl(\frac{\sg_y(r)}{v}\biggr)\,\ga'\Bigl(\frac{r}{y}\Bigr)\,
e^{-2\pi i(x/y - u/v)(r - \sg_y(r))}\,\dl(\sg_v(\sg_y(r)) - r)\,dr.
\end{align*}
The argument of this last delta vanishes if and only if
$\sg_v(r) = \sg_y(r)$, that is, if and only if $v = y$. On setting
$f(v) := \sg_v(\sg_y(r)) - r$, it follows that
$$
f'(v)\bigr|_{v=y} = \sg\Bigl( \sg\Bigl(\frac{r}{y}\Bigr) \Bigr)
- \sg'\Bigl( \sg\Bigl(\frac{r}{y}\Bigr) \Bigr)
\,\sg\Bigl(\frac{r}{y}\Bigr)
= \frac{r}{y} - \sg'\Bigl( \sg\Bigl(\frac{r}{y}\Bigr) \Bigr)
\,\sg\Bigl(\frac{r}{y}\Bigr)
= \frac{r}{y} \,\sg\Bigl(\frac{r}{y}\Bigr)
\,\ga'\Bigl( \sg\Bigl(\frac{r}{y}\Bigr) \Bigr),
$$
using \eqref{eq:sg-trick}. Therefore,
$$
\frac{r}{y} \,\sg\Bigl(\frac{r}{y}\Bigr)\,
\ga'\Bigl(\sg\Bigl(\frac{r}{y}\Bigr)\Bigr)\,\dl(r - \sg_v(\sg_y(r)))
= \dl(v - y),
$$
and~\eqref{eq:omi-trace} follows at once:
\begin{equation*}
\Tr\bigl(\Om(w)\Om(z)\bigr)
= \dl(v - y) \int_0^\infty
\ga'\Bigl(\frac{r}{y}\Bigr)\, e^{-2\pi i(x - u) \ga(r/y)} \,dr
= y \,\dl(x - u) \,\dl(v - y).
\tag*{\qed}
\end{equation*}
\hideqed
\end{proof}

\begin{corl}
\label{cr:tracial}
The maps \eqref{eq:vice-versa} establish an isometric
isomorphism between the Hilbert space of Hilbert--Schmidt operators
on~$\K$ and the Hilbert space $L^2(\Pi,d\om(z))$ of square-summable
functions on the upper Poincar\'e half-plane with the left-invariant
measure.
\qed
\end{corl}

\subsection{Relation with the Wigner functions on the half-plane}
\label{ssc:Wigner}

Starting from~\eqref{eq:omz-ker-joe}, by dequantization we obtain
\begin{align}
W_A(z) &= \Tr(\Om(z)A) = \int_0^\infty \int_0^\infty
A(r,s)\,\Om(z;s,r) \,\frac{dr}r \,\frac{ds}s
\nonumber \\
&= \frac{1}{y} \int_0^\infty \int_0^\infty
A(r,s)\,\ga'(s/y)\,e^{2\pi ix\ga(s/y)} \,\dl(r-\sg_y(s)) \,dr\,ds
\nonumber \\
&= \frac{1}{y} \int_0^\infty A(\sg_y(s), s) \,\ga'(s/y)\,
e^{2\pi ix\ga(s/y)} \,ds
\nonumber \\
&= \int_{-\infty}^\infty A(y\la(u), y\la(-u))\, e^{-2\pi ixu} \,du.
\label{eq:dequant}
\end{align}
Note how by means of \eqref{eq:omz-ker-vic} we recover the
kernel of~$A$ from its dequantization:
\begin{equation}
A(r,s) = \int_\Pi W_A(z) \Om(z;r,s) \,\frac{dx\,dy}{y}
= \int_{-\infty}^\infty W_A\biggl(x, \frac{r - s}{\log(r/s)} \biggr)
\,e^{2\pi ix\log(r/s)} \,dx.
\label{eq:quant-kers}
\end{equation}

In particular, the Wigner function $W^\psi$ of a state $\psi$ is
simply the expected value of the Stratonovich--Weyl operator~$\Om$ or
the dequantization of the projector $\ketbra{\psi}{\psi}$:
$$
W^\psi(z) := \braket{\psi}{\Om(z)\psi} = \braket{\Om(z)\psi}{\psi}
= \Tr\bigl(\Om(z)\,\ketbra{\psi}{\psi}\bigr).
$$
The operator $\ketbra{\psi}{\psi}$ has kernel
$A(r,s) = \psi(r)\psi^*(s)$. Thus,
\begin{equation}
W^\psi(z) = \int_\R e^{-2\pi ixu}\,\psi(y\la(u))\,\psi^*(y\la(-u))\,du.
\label{eq:Wigner}
\end{equation}
Now, for the upper half-plane, a plethora of ``affine Wigner
functions'' were originally constructed by the Bertrands
\cite{BertrandsAff}; see also~\cite{MolnarBB}. One family among
several options satisfying good covariance properties (under the
`$ax+b$' group and some extensions of it) is distinguished by
``unitarity'', that is, the correspondence
$\ketbra{\psi}{\psi} \to W^\psi$ should extend to a unitary
isomorphism between Hilbert--Schmidt operators and
$L^2(\Pi,dx\,dy/y)$. For our purposes, it is enough to check that
\eqref{eq:Wigner} with our quantizer coincides with their Wigner
functions. This is done by inspection of formula (57)
in~\cite{MolnarBB}, modulo our conventions for the unirrep~$K[\Oh_+]$,
or equation~(IV.7) in~\cite{BertrandsSymb}, where Darboux coordinates
on the phase space are used.

For an arbitrary normalized state~$\psi$, we remark that
$\int_\Pi |W^\psi(z)|^2 \,d\om(z) = 1$, which seems curious since
$\psi$ need not belong to the domains of all~$\Om(z)$. We return to
this question in~\cite{Eris}. Geometrical properties of the affine
Wigner functions have been much investigated, in regard to positivity,
localization, marginal distributions, interference, etc. On this, we
can do little better than to refer to Flandrin's
articles~\cite{Flandrin}.

\subsection{Summary}
\label{ssc:pelota}

The strategy outlined in subsection~\ref{ssc:Weyl} has been
successful. The outcome is that the deceptively simple formula for the
operator-valued distribution
\begin{equation}
\D(\Pi) \hatox \B(\K) \ni \Om(F)
:= \int_\aff e^{-2\pi i\<F,X>} U(\exp X)\,dX,
\label{eq:etsi-non-daretur}
\end{equation}
with $dX$ the Lebesgue measure, makes sense, and the bounded operators
$$
\int_{\aff^*} a(F) \int_\aff e^{-2\pi i\<F,X>} U(\exp  X) \,dX\,dF
$$
for $a$ a test function on~$\aff^*$, supported on~$\Pi$, are
explicitly given. Lest the reader be misled by the heuristic approach
ostensibly taken in subsection~\ref{ssc:Weyl}, it must be said
that~\eqref{eq:etsi-non-daretur} recommends itself because from it
covariance of~$\Om$ is ensured to hold by an abstract argument. Again,
the reader will have no difficulty in checking it. This point of view
had been emphasized in~\cite{UnterbergerPrivee}.

\section{The Moyal twisted product on the half-plane}
\label{sec:product}

The \textit{Moyal product} $f\star h$ of two functions $f,h$ on~$\Pi$
is by definition the dequantization of the operator $Q(f)Q(h)$; to
wit,
\begin{align}
f \star h(z) &:= \Tr\biggl[ \Om(z) \int f(w) \Om(w) \,d\om(w)
\int g(t) \Om(t) \,d\om(t) \biggr]
\nonumber \\
&= \iint_{\Pi^2} K_\star(z,w,t) f(w) h(t) \,d\om(w) \,d\om(t),
\label{eq:el-remate}
\end{align}
where
\begin{equation}
K_\star(z,w,t) := \Tr\bigl( \Om(z)\Om(w)\Om(t) \bigr).
\label{eq:run-around}
\end{equation}
The ``trikernel'' $K_\star$ enjoys left invariance:
$$
\Tr\bigl( \Om(Z\.z)\Om(Z\.w)\Om(Z\.t) \bigr)
= \Tr\bigl( U(g_Z) \Om(z)\Om(w)\Om(t) U^\7(g_Z) \bigr)
= \Tr\bigl( \Om(z)\Om(w)\Om(t) \bigr).
$$
Using this invariance we can rewrite the Moyal product of two
functions ---also, using \eqref{eq:alpha-omega} and the cyclicity of
the trace-integral, eventually of many distributions~\cite{Phobos}---
on~$\Pi$, gifted with the invariant measure $d\om(z) = dx\,dy/y$, in
the following ways:
\begin{align*}
f \star h(z) &= \iint_{\Pi^2} K_\star(z,w,t) f(w)
h(t)\,d\om(w)\,d\d\om(t)
\\
&= \iint_{\Pi^2} K_\star(i,z^{-1}\.w,z^{-1}\.t) f(w) h(t)
\,d\om(w)\,d\om(t)
\\
&= \iint_{\Pi^2} K_\star(i,w,t) f(z\.w) h(z\.t)\,d\om(w)\,d\om(t).
\end{align*}
Let $R_w$, $R_t$ denote right multiplication operators (i.e., the
right regular action) for the group structure on~$\Pi$, and note the
elegance of the final ``deformation'' formula:
\begin{equation}
f \star h
= \iint_{\Pi^2} K_\star(i,w,t) \,R_wf \,R_th \,d\om(w)\,d\om(t).
\label{eq:left-of-right}
\end{equation}

We do not omit, finally, the \textit{tracial identity} for our star
product:
$$
\int_\Pi f \star h(z) \,d\om(z) = \int_\Pi f(z) h(z) \,d\om(z);
$$
this comes straight from~\eqref{eq:el-remate} on using the properties
of the Stratonovich--Weyl quantizer.

\subsection{The trikernel for the twisted product and its symmetries}
\label{ssc:tri-ker}

We need the solution $\ka(y_0,y_1,y_2)$ of the equation
$$
s = \sg_{y_0}(\sg_{y_1}(\sg_{y_2}(s))).
$$
Equivalently, it is the solution of
$$
\sg_{y_0}(s) = \sg_{y_1}(\sg_{y_2}(s)), \word{or}
\sg_{y_2}(s) = \sg_{y_1}(\sg_{y_0}(s)).
$$
Given any (positive) values of $y_0,y_1,y_2$, there is indeed a unique
solution to these equations, for one of the sides increases
monotonically from $0$ to~$\infty$, whereas the other decreases from
$\infty$ to~$0$. Under the exchange $z_0 \otto z_2$, with $z_1$ held
fixed, the value of~$\ka$ is unchanged. Moreover, $\ka$ is a
homogeneous function of degree~$1$, since
$c^{-1}\,\ka(cy_0,cy_1,cy_2)$ and $\ka(y_0,y_1,y_2)$ solve the same
equation. We abbreviate
$\sg_{012} := \sg_{y_0} \circ \sg_{y_1} \circ \sg_{y_2}$, whose
inverse function is
$\sg_{210} := \sg_{y_2} \circ \sg_{y_1} \circ \sg_{y_0}$.

With this in hand, one can proceed to compute the trikernel. After a
straightforward, though tedious, calculation, one obtains
\begin{align}
& K_\star(z_0,z_1,z_2) 
\nonumber \\
&= \frac{\ka\, \sg_{y_0}(\ka)\, \sg_{y_2}(\ka)\, \ga'_{y_2}(\ka)\,
\ga'_{y_0}(\sg_{y_0}(\ka))\, \ga'_{y_1}(\sg_{y_2}(\ka))}
{y_0y_1y_2\,\bigl(1 - \sg'_{012}(\ka)\bigr)} \,\,
e^{2\pi i\bigl[\frac{x_0}{y_0}\ga_{y_0}(\ka) - \frac{x_1}{y_1}
\ga_{y_1}(\sg_{y_2}(\ka)) - \frac{x_2}{y_2} \ga_{y_2}(\ka) \bigr]}
\label{eq:tri-ker-overt}
\end{align}
where the dependence on $y_0,y_1,y_2$ through $\ka$ is understood.

The trikernel is of the general form
$$
K_\star(z_0,z_1,z_2) = A(z_0,z_1,z_2)\,e^{2\pi iS(z_0,z_1,z_2)}
$$
for real amplitude $A\: M^3 \to \R$ and phase $S\: M^3 \to \R$
functions. By its construction, we expect $K_\star$ to have several
symmetries. First of all, cyclical symmetry. With
$\hat\ka(y_0,y_1,y_2)$ defined by $\hat\ka = \sg_{201}(\hat\ka)$, with
the obvious notation, the same calculation gives
\begin{align}
& K_\star(z_2,z_0,z_1) 
\nonumber \\
&= \frac{\hat\ka\, \sg_{y_2}(\hat\ka)\, \sg_{y_1}(\hat\ka)\,
\ga'_{y_1}(\hat\ka)\, \ga'_{y_2}(\sg_{y_2}(\hat\ka))\,
\ga'_{y_0}(\sg_{y_1}(\hat\ka))}
{y_0y_1y_2\, \bigl(1 - \sg'_{201}(\hat\ka)\bigr)} \,\,
e^{2\pi i\bigl[ \frac{x_2}{y_2}\ga_{y_2}(\hat\ka)
- \frac{x_0}{y_0} \ga_{y_0}(\sg_{y_1}(\hat\ka))
- \frac{x_1}{y_1}\ga_{y_1}(\hat\ka) \bigr]}.
\label{eq:tri-ker-spun}
\end{align}
But $\hat\ka$ is just $\sg_{y_2}(\ka)$; thus
$$
\ga_{y_2}(\ka) = - \ga_{y_2}(\hat\ka), \word{and also}
\sg_{y_0}(\ka) = \sg_{y_1}(\hat\ka), \word{implying}
\ga_{y_0}(\sg_{y_1}(\hat\ka)) = - \ga_{y_0}(\ka),
$$
and one sees at once that the numerator of the fraction in
\eqref{eq:tri-ker-overt} and the phase factor coincide with those of
the new formula \eqref{eq:tri-ker-spun}. Moreover,
$$
\sg'_{201}(\hat\ka) 
= \sg'_{y_2}(\sg_{y_0}(\sg_{y_1}(\hat\ka)))\,
\sg'_{y_0}(\sg_{y_1}(\hat\ka))\, \sg'_{y_1}(\hat\ka)
= \sg'_{y_2}(\ka)\, \sg'_{y_0}(\sg_{y_1}(\sg_{y_2}(\ka)))\,
\sg'_{y_1}(\sg_{y_2}(\ka)) = \sg'_{012}(\ka),
$$
and we conclude that, as expected,
$$
K_\star(z_0,z_1,z_2) = K_\star(z_2,z_0,z_1) = K_\star(z_1,z_2,z_0).
$$
Next we investigate the switch $z_0 \otto z_2$ (with $z_1$ fixed). We
observe that
$$
\frac{1}{1 - \sg'_{012}(\ka)} = \frac{-
\sg'_{210}(\sg_{210}(\ka))}{1 - \sg'_{210}(\sg_{210}(\ka))} =
\frac{- \sg'_{210}(\ka)}{1 - \sg'_{210}(\ka)},
$$
since $\sg_{210}(\ka)=\ka$. On multiplying the numerator of the
fraction occurring in~\eqref{eq:tri-ker-overt} by
$$
- \sg'_{210}(\ka) = - \sg'_{y_2}(\sg_{y_1}(\sg_{y_0}(\ka)))
\,\sg'_{y_1}(\sg_{y_0}(\ka)) \,\sg'_{y_0}(\ka)
$$
and taking \eqref{eq:gay-sgy-relns} into account, the whole fraction
becomes
$$
\frac{\ka \,\sg_{y_0}(\ka)\,\sg_{y_2}(\ka)
\,\ga'_{y_2}((\sg_{y_2}(\ka)) \,\ga'_{y_1}(\sg_{y_0}(\ka))
\,\ga'_{y_0}(\ka)} {y_0y_1y_2\,(1 - \sg'_{210}(\ka))}.
$$
In other words, the fraction is unchanged by the switch
$z_0 \otto z_2$. Now note that
\begin{equation*}
\ga_{y_1}(\sg_{y_2}(\ka))
= \sg_{y_2}(\ka) - \sg_{y_1}(\sg_{y_2}(\ka))
= \sg_{y_2}(\ka) - \sg_{y_0}(\ka)
= \ga_{y_0}(\ka) - \ga_{y_2}(\ka).
\end{equation*}
Using this formula, we can reexpress the phase factor in the trikernel
\eqref{eq:tri-ker-overt} as
$$
\exp\biggl\{ 2\pi i \biggl[
\Bigl( \frac{x_0}{y_0} - \frac{x_1}{y_1} \Bigr) \,\ga_{y_0}(\ka) +
\Bigl( \frac{x_1}{y_1} - \frac{x_2}{y_2} \Bigr) \,\ga_{y_2}(\ka)
\biggr] \biggr\}.
$$
This is manifestly \textit{skewsymmetric} under the exchange
$z_0 \otto z_2$, with $z_1$ held fixed. In fine, we have shown that
$$
\owl{K_\star(z_0,z_1,z_2)} = K_\star(z_2,z_1,z_0) =
K_\star(z_0,z_2,z_1) = K_\star(z_1,z_0,z_2),
$$
where cyclic symmetry has been reinvoked; in particular, this confirms
that complex conjugation is an antilinear involution for the twisted
product. Corollary~\ref{cr:tracial} can now be read as stating that
$(L^2(\Pi,d\om),\star)$ is a Hilbert algebra.

Finally, we expect
$K_\star(z_0,z_1,z_2) = K_\star(i,z_0^{-1}\.z_1,z_0^{-1}\.z_2)$, in
view of left invariance. Indeed, the trikernel is invariant under the
transformations
$$
\begin{aligned}
x_0 &\mapsto 0, \\ y_0 &\mapsto 1,
\end{aligned} \qquad
\begin{aligned}
x_1 &\mapsto x_1 - x_0y_1/y_0, \\ y_1 &\mapsto y_1/y_0, 
\end{aligned} \qquad
\begin{aligned}
x_2 &\mapsto x_2 - x_0y_2/y_0, \\ y_2 &\mapsto y_2/y_0,
\end{aligned}
$$
on account of the homogeneity properties of~$\ka$.

Inspired by earlier exact results~\cite{Iapetus,WildbergerFou},
Weinstein~\cite{WeinsteinTri} developed a heuristic argument for the
construction of trikernels on symplectic symmetric spaces. In this
approach, the phase function~$S$ is postulated to be an (invariant)
oriented symplectic area of a geodesic triangle for which
$z_0,z_1,z_2$ are the midpoints of the sides, and the amplitude~$A$ is
chosen as to achieve associativity of the twisted product ---implicit
in our treatment--- and other desirable properties. The idea has been
further developed in \cite{RiosT} ---where some caveats are made---
and in \cite{Bieliavsky} and \cite[\S\,3.3.5]{Detournay}. It is
sometimes linked to the purported role of reflections in producing
quantizers. However, it is known~\cite{Unterbergers} that reflections
do not lead directly to Stratonovich--Weyl quantizers in general; and
it is straightforward to verify that, although it enjoys the same
symmetries, our phase function is not the area. The question deserves
further investigation~\cite{BieliavskyGI}.

\subsection{The extended covariance group of the twisted product}
\label{ssc:exten-symm}

The ordinary Moyal product on the full plane $\R^2$ has a larger
covariance group than the original Heisenberg group of phase-space
translations under which it is equivariant; this is the inhomogeneous
metaplectic group of unitaries $U(g)$ such that
$Q(f) \mapsto U(g)\,Q(f)\,U^\7(g) =: Q(f \circ \vf)$ implements a
diffeomorphism $\vf$ of the plane that normalizes the action of the
Heisenberg group. For the half-plane $\Pi$, its analogue will be a Lie
group of symplectomorphisms normalizing the action of $\Aff$. At the
infinitesimal level, the generators of this group are given by symbols
$f_i$ such that the Moyal bracket
$$
[f_i, h]_\star := 2\pi i(f_i \star h - h \star f_i)
$$
coincides, for arbitrary $h$, with the Poisson bracket
$$
\{f_i, h\}_\PB = y \biggl( \pd{f_i}{x} \pd{h}{y}
- \pd{f_i}{y} \pd{h}{x} \biggr)
$$
corresponding to the symplectic $2$-form $dx \w dy/y$ on~$\Pi$. In 
other words, these $f_i$ are ``distinguished observables'' in the 
sense of~\cite{BayenFFLS}.

The (neutral component of) the normalizer of $\Aff$ within the group
of symplectomorphisms of~$\Pi$ is easily determined
\cite{UnterbergerPrivee}. Any one-parameter subgroup is generated by 
a Hamiltonian vector field of the form
$H_f = y\bigl( f_y\,\pd{}{x} - f_x\,\pd{}{y} \bigr)$ for some
$f \in \Coo(\Pi)$. Since the action of $\aff$ on~$\Pi$ is generated by
the vector fields $y\,\pd{}{x}$ and $y\,\pd{}{y}$, we require that
$$
\biggl[ H_f, y\,\pd{}{x} \biggr]
= -(f_{xy} + f_y) y\,\pd{}{x} + y^2 f_{xx}\,\pd{}{y},  \qquad
\biggl[ H_f, y\,\pd{}{y} \biggr]
= -(y f_{yy} + f_y) y\,\pd{}{x} + y^2 f_{xy}\,\pd{}{y}
$$
be linear combinations of $y\,\pd{}{x}$ and $y\,\pd{}{y}$. This easily
entails that 
$$
f(x,y) = \al x + \bt y + \ga\log y + \dl
$$
for some constants $\al,\bt,\ga,\dl$. Ignoring the trivial constant
term that does not contribute to
$H_f = -\al y\,\pd{}{y} + \bt y\,\pd{}{x} + \ga\,\pd{}{x}$, we obtain
a solvable $3$-parameter group $G$ extending $\Aff$ by~$\R$. The
appearance of the $\log$ function above is related to the existence of
ray unirreps of~$G$ given by
$$
U(a,b,c) \psi(r) = e^{2\pi ibr} r^{2\pi ic} \psi(ar).
$$
This covariance group was also found in~\cite{BertrandsAff} by a not
very different method.

To ascertain that this group is indeed a symmetry group of our twisted
product, one must verify that the three functions $x$, $y$, $\log y$
are distinguished observables.

\begin{lem}
\label{lm:star-comm}
For any smooth function $h$ on~$\Pi$, the following relations hold:
\begin{equation}
[x, h]_\star = y \pd{h}{y};  \qquad
[y, h]_\star = - y \pd{h}{x};  \qquad
[\log y, h]_\star = - \pd{h}{x}.
\label{eq:star-comm}
\end{equation}
\end{lem}

\begin{proof}
We first determine the operator kernels corresponding,
via~\eqref{eq:quant-kers}, to the three basic functions:
\begin{align}
Q_x(r,s) &= \int_\R x \,e^{2\pi ix\log(r/s)} \,dx
= \frac{1}{2\pi i} \dl'(\log r - \log s)
= \frac{1}{2\pi i} \bigl( s^2\,\dl'(r - s) - s\,\dl(r - s) \bigr),
\nonumber \\
Q_y(r,s)
&= \int_\R \frac{r - s}{\log(r/s)} \,e^{2\pi ix\log(r/s)} \,dx
= \frac{r}{\la(\log(r/s))} \,\dl(\log r - \log s) 
= \frac{rs}{\la(\log(r/s))} \,\dl(r - s),
\nonumber \\
Q_{\log y}(r,s)
&= \int_\R \log\biggl( \frac{r - s}{\log(r/s)} \biggr)
\,e^{2\pi ix\log(r/s)} \,dx
= \frac{r}{\la(\log(r/s))} \,\dl(\log r - \log s) 
\nonumber \\
&= \bigl( s\log r - s \log\la(\log(r/s)) \bigr) \,\dl(r - s).
\label{eq:nice-kernels}
\end{align}
We have written $Q_x$ for $Q(x)$, and similarly for the other
operators. If the quantized operator $Q(h)$ has kernel $B(s,t)$, then
$Q\bigl([x,h]_\star\bigr) = 2\pi i[Q_x, Q(h)]$ has kernel
\begin{align*}
2\pi i & \int_0^\infty
\bigl( Q_x(r,t)\,B(t,s) - B(r,t)\,Q_x(t,s)  \bigr) \,\frac{dt}{t}
\\
&= \int_0^\infty \bigl( t\,\dl'(r - t) - \dl(r - t) \bigr) B(t,s)
- B(r,t) \bigl( t\,\dl'(t - s) - \dl(t - s) \bigr) \,dt
\\
&= \pd{}{t}\biggr|_{t=r}(-t B(t,s)) - \pd{}{t}\biggr|_{t=s}(-t B(r,t))
= -r \pd{B}{r}(r,s) + s\,\pd{B}{s}(r,s).
\end{align*}
On the other hand, \eqref{eq:dequant} yields
\begin{align*}
y\,\pd{h}{y}(z) 
&= \int_\R \biggl( y\la(u) \,\pd{B}{r} \bigl(y\la(u), y\la(-u)\bigr)
+ y\la(-u) \,\pd{B}{s} \bigl(y\la(u), y\la(-u)\bigl) \biggr)
\,e^{-2\pi ixu} \,du
\\
&= \frac{1}{y} \int_0^\infty r\,\pd{B}{r}(r, \sg_y(r)) \,\ga'_y(r)
\,e^{-2\pi ix\ga(r/y)} \,dr
+ \frac{1}{y} \int_0^\infty s\,\pd{B}{s}(\sg_y(s), s) \,\ga'_y(s)
\,e^{2\pi ix\ga(s/y)} \,ds
\\
&= \frac{1}{y} \int_0^\infty \int_0^\infty 
\biggl( -r\,\pd{B}{r} + s\,\pd{B}{s} \biggr)(r,s) \,\ga'_y(s)
e^{2\pi ix\ga(s/y)} \,\dl(r - \sg_y(s)) \,dr\,ds,
\end{align*}
and using \eqref{eq:dequant} once more we obtain the desired relation:
$$
\{x, h\}_\PB = y\,\pd{h}{y} = W_{Q([x,h]_\star)} = [x,h]_\star.
$$

The other cases are simpler. One finds from~\eqref{eq:nice-kernels}
that $Q([y,h]_\star) = 2\pi i [Q_y, Q(h)]$ has kernel
$2\pi i(r - s)\,B(r,s)$ and that $Q([\log y,h]_\star)$ has kernel
$2\pi i (\log r - \log s)\,B(r,s)$. Therefore,
\begin{align*}
[y, h]_\star(z) 
&= \frac{2\pi i}{y} \int_0^\infty (\sg_y(s) - s) \,B(\sg_y(s), s)
\,\ga'_y(s) \,e^{2\pi ix\ga(s/y)} \,ds
\\
&= - 2\pi i \int_0^\infty B(\sg_y(s), s) \,\ga'_y(s)
\bigl( \ga(s/y) \,e^{2\pi ix\ga(s/y)} \bigr) \,ds
\\
&= - \pd{}{x} \int_0^\infty B(\sg_y(s), s) \,\ga'_y(s)
\,e^{2\pi ix\ga(s/y)} \,ds
\\
&= - \pd{}{x} (y h(z)) = -y \pd{h}{x}(z).
\end{align*}
An almost identical calculation, with the relation
$\sg_y(s) - s = - y\,\ga(s/y)$ replaced by the identity
$\log \sg_y(s) - \log s = - \ga(s/y)$, yields 
$[\log y, h]_\star = - \del h/\del x$. Or we may just remark that
$[\cdot, h]_\star$ is a derivation. 
\end{proof}

On regarding the functions $x,y$ on phase space as elements of the Lie
algebra~$\aff$, the first two equalities of~\eqref{eq:star-comm} show
that our twisted product is an $\aff$-invariant $\star$-quantization
in the sense of~\cite{BayenFFLS,Fronsdal}.

\subsection{On the universal enveloping algebra product}
\label{ssc:univ-rollo}

By duality and symmetrization, the universal enveloping algebra
$\U(\g)$ of a Lie algebra~$\g$ can be realized by an algebra
$\P(\g^*)$ of polynomial functions on~$\g^*$. Thus it makes sense to
compare the (restriction to~$\Pi$ of) the product on~$\U(\aff)$ as
transferred to~$\P(\g^*)$ with the twisted product; in our case the
correspondence gives $X_1 \mapsto x \equiv x_1$;
$X_2 \mapsto y \equiv x_2$. Let us denote by~$*$ that transported
product. Its expression is complicated in general, but it is well
known that one obtains
\begin{align}
x_i * h &= x_i h + \sum_{n=1}^\infty B_n \sum_{k_1\dots k_n}
\{\ad(x_{k_1}) \dots \ad(x_{k_n})\}_\sym(x_i)
\frac{\del^n h} {\del x_{k_1}\dots\del x_{k_n}};
\label{eq:apocalypse-now} \\
h * x_i &= x_i h + \frac{1}{2} \sum_j [x_j,x_i] \pd{h}{x_j}
+ \sum_{n=2}^\infty B_n \sum_{k_1\dots k_n}
\{\ad(x_{k_1}) \dots \ad(x_{k_n})\}_\sym(x_i)
\frac{\del^n h} {\del x_{k_1}\dots\del x_{k_n}};
\nonumber
\end{align}
where the $k_j$ take the values~$1$ or~$2$ in all possible forms and
$\{\dots\}_\sym$ means total symmetrization of the operations inside
the curly brackets. The rule makes sense because the Lie products are
supposed known: here $[x_1,x_2] = x_2$. For instance, we see that, if
$h$ depends only on the second variable, then the series terminates,
and $x * h(x,y) = x h(y) + \half y h'(y)$. The second term of the
series in~\eqref{eq:apocalypse-now} is just $\half$ times the Poisson
bracket:
$$
\sum_j [x_j,x] \pd{h}{x_j} = -y\pd{h}{y}; \qquad
\sum_j [x_j,y] \pd{h}{x_j} =  y\pd{h}{x}.
$$
Thus in particular
$$
x_i * h - h * x_i = \{x,h\}_\PB = [x_i, h]_\star.
$$
A detailed comparison between $*$ and the asymptotic expansion
of~$\star$ would lengthen this paper too much; we come back on this
matter in \cite{Eris} and~\cite{BieliavskyGI}.

\section{Right-covariant quantization}
\label{sec:righteous}

In this section the notation $d_lz$ will be used for
$d\om(z) = dx\,dy/y$, and $d_rz$ for $dx\,dy/y^2$. For the purposes
outlined in the introduction, we are actually interested in a
right-covariant star product, as well as the left-covariant one
constructed so far. We summarize again our \textit{desideratum}: a
pair of quantizers $\Om^L$, $\Om^R$, both acting on~$\K$, satisfying
\begin{enumerate}
\item[(i)]
$\Om^{L,R}(z)^\7 = \Om^{L,R}(z)$,
\item[(ii)]
$U(g_{z'})\,\Om^L(z)\,U^\7(g_{z'}) = \Om^L(z'\.z), \word{and}
U^\7(g_{z'})\,\Om^R(z)\,U(g_{z'}) = \Om^R(z\.z')$,
\item[(iii)]
$\Tr \Om^{L,R}(z) = 1$,
\item[(iv)]
$\Tr\bigl( \Om^{L,R}(z)\,\Om^{L,R}(z') \bigr) = I_{L,R}(z,z')$,
\end{enumerate}
with $I_L$ and $I_R$ denoting the reproducing kernels for
$L^2(\Pi, d_l z)$ and $L^2(\Pi, d_r z)$, respectively. Here
$\Om^L \equiv \Om$ is the quantizer already found. Define
$$
\breve f(z) := f(z^{-1}), \word{with}
z^{-1} = (x + iy)^{-1} := -x/y + i/y,
$$
the inverse for the product~\eqref{eq:orbit-prod}. It seems natural to
replace the quantization rule~\eqref{eq:vice-versa} by
$$
A = \int_\Pi \Om^R(z)\,f(z) \,d_rz =: Q_R(f),
$$
when using the right-covariant quantizer. We obtain
$Q_R(\breve f) = Q(f)$ if we declare that
$$
\Om^R(z) \equiv \Om(z^{-1}).
$$

It is not idle to check consistency of this rule with previous use of
the coadjoint action and the diffeomorphism $z \mapsto g_z$. We need
to verify that
\begin{equation}
z \rt g_{z'} \equiv (g_{z'}^{-1} \lt z^{-1})^{-1} = z\.z'.
\label{eq:right-coad}
\end{equation}
Indeed,
$$
\bigl( (y',-x') \lt (-x/y + i/y) \bigr)^{-1}
= \biggl(\frac{-x'-xy'}{yy'} + \frac{i}{yy'} \biggr)^{-1}
= x' + xy' + iyy'.
$$
The right-covariant quantizer is thus given by the following
expression:
$$
\Om^R(z) = \frac{1}{y} \iint_{\R^2} e^{2\pi i(ux/y - v/y)}\,
U(e^u, v/\la(-u)) \,du\,dv.
$$
It is easy to check consistency of this with $\Om(z) = \Om^R(z^{-1})$.
It is also straightforward to verify, along the same lines as before,
the four properties listed above; in particular,
$$
\Tr\bigl( \Om^R(z)\,\Om^R(w) \bigr) = y^2\,\dl(z - w).
$$
For the trikernel, one obtains
$$
K^R_\star(z,w,t) = \Tr\bigl( \Om^R(z)\,\Om^R(w)\,\Om^R(t) \bigr)
= K_\star(z^{-1},w^{-1},t^{-1}).
$$
This yields the following tautological relation between the twisted
products $\star$ and~$\star^R$\,:
\begin{align*}
f\star^R h(z)
&= \int_{\Pi^2} K^R_\star(z,w,t)\,f(w)\,h(t) \,d_rw\,d_rt
\\
&= \int_{\Pi^2} K_\star(z^{-1},w^{-1},t^{-1})\,f(w)\,h(t) \,d_rw\,d_rt
\\
&= \int_{\Pi^2} K_\star(z^{-1},w,t)\,f(w^{-1})\,h(t^{-1}) \,d_lw\,d_lt
= (\breve f \star \breve h)\breve{\ }(z),
\end{align*}
consistent with
$$
Q_R(f \star^R h) = Q_R(f)\,Q_R(h).
$$

We finally register the following formula for the product $\star^R$,
similar to~\eqref{eq:left-of-right}:
$$
f \star^R h = \iint_{\Pi^2} K_\star(i,w,t) L_w f L_t h \,d_lw \,d_lt,
$$
and the tracial identity for the $\star^R$ product:
$$
\int_\Pi f \star^R h(z) \, d_rz = \int_\Pi f(z) h(z) \,d_rz.
$$

\section{Fourier--Moyal transformations on the `$ax + b$' group}
\label{sec:FM-Aff}

\subsection{The Fourier--Moyal kernels}
\label{ssc:FM-ker}

Consider the following distribution or kernel:
$$
\EE(z,g) \equiv \EE_L(z,g) := \Tr\bigl( \Om_\pm(z)\,U_\pm(g) \bigr)
= W_{U_\pm(g)}(z),
$$
for $z \in \pm\Pi$, respectively, and $g \in \Aff$. As a consequence
of this definition, we see that $\EE$ is the symbol for the unirreps:
\begin{equation}
U_\pm(g) = \int_{\pm\Pi} \EE(z,g) \,\Om(z) \,d\om_\pm(z).
\label{eq:rep-symbol}
\end{equation}

We compute the kernel, expecting the covariance property
\begin{equation}
\EE(h \lt z, hgh^{-1}) = \EE(z, g).
\label{eq:FM-covar}
\end{equation}
It comes as a ``nice surprise'' that the kernel is a $U(1)$-valued
smooth function. Take $\Im z > 0$. From equation
\eqref{eq:Weyl-Ansatz}, one gets
$$
\Omega(z)\,U(g) = y \iint_{\R^2} e^{-2\pi i(xu + yv)}\,
U(e^ua, e^u(b + v/\la(u)) \,du\,dv,
$$
and the kernel of this operator is thus
\begin{align}
\Om U(z,g;r,s)
&= y \iint_{\R^2} e^{-2\pi i(xu + yv)}\,
e^{2\pi i e^u(b + v/\la(u))r}\, s\,\dl(s - re^ua) \,du\,dv
\nonumber \\
&= y \iint_{\R^2} e^{-2\pi i(xu + yv)}\, e^{2\pi ie^u(b + v/\la(u))r}
\,\dl(u - \log(s/ra)) \,du\,dv
\nonumber \\
&= y \int_\R e^{-2\pi i(x\log(s/ra) + yv)}\,
e^{2\pi ia^{-1}s(b + v/\la(\log(s/ra)))} \,dv
\nonumber \\
&= y\,e^{-2\pi ix\log(s/ra)}\, e^{2\pi ibs/a}
\,\dl\bigl( y - s/a\,\la(\log(s/ra)) \bigr).
\label{eq:ker-OmU}
\end{align}
Its trace is the desired kernel, explicitly:
\begin{align}
\EE(z,g) &= \int_0^\infty\,\Om U(z,g;r,r) \,\frac{dr}{r}
\nonumber \\
&= y\,e^{2\pi ix\log a} \int_0^\infty e^{2\pi irb/a}
\,\dl\bigl(y - r/\la(\log a)\bigr) \,\frac{dr}{r}
\nonumber \\
&= y\,e^{2\pi ix\log a} \int_0^\infty e^{2\pi irb/a}\,
\la(\log a) \,\dl\bigl(r - y\la(\log a)\bigr) \,\frac{dr}{r}
\nonumber \\
&= e^{2\pi i(x\log a + y\,b\la(\log a)/a)}.
\label{eq:trial-by-fire}
\end{align}

We then see that $\EE$ is smooth and of modulus~$1$. We check the
coadjoint covariance of this kernel. With $h = (a',b')$, one indeed
finds that
$$
\EE(x + b'y/a' + iy/a'; (a, a'b + b'(1-a))) = \EE(x + iy;(a,b)),
$$
since $(a^{-1} - 1)\la(\log a) = -\log a$ from the definition
\eqref{eq:lambda-curve}. By computing on the second orbit, the formula
is valid for $y < 0$ as well.

Similarly, the right-covariant quantizer yields another kernel:
$$
\EE_R(z,g) := \Tr\bigl( \Om^R(z)\,U(g) \bigr) = \EE(z^{-1},g),
$$
with the expected covariance property:
$$
\EE_R(z \rt h, h^{-1}gh) = \EE_R(z, g),
$$
which follows from \eqref{eq:FM-covar}, in view
of~\eqref{eq:right-coad}. Thus we get, explicitly,
$$
\EE_R(z,g) = e^{2\pi i(-x\log a + b\la(\log a)/a)/y}.
$$

It is enlightening to pass to Lie-algebra coordinates in the
$\EE$-functions, for the first group arguments:
$$
\EE(z;u,b) = e^{2\pi i( ux + b\la(-u)y)},  \qquad
\EE_R(z;u,b) = e^{2\pi i(-ux + b\la(-u))/y};
$$
or for both:
$$
\EE(z;u,v) = e^{2\pi i(ux + vy)},  \qquad
\EE_R(z;u,v) = e^{2\pi i(-ux/y + v/y)}.
$$
Note the simplicity of the last result for~$\EE$. If we regard $z$ as
an element of~$\aff^*$, then
$$
\Tr\bigl( \Om_\pm(z)\,U_\pm(\exp X) \bigr)
= \exp\bigl( 2\pi i \<z,X> \bigr),  \word{for all}  X \in \aff,
$$
has been proved to hold.

In conclusion, the \textit{left Fourier--Moyal transformation},
denoted $\FM$, is given by
\begin{align*}
\FM[f](z) &:= \int \EE(z,g)\, f(g) \,d_lg
\\
&= \iint_{\aff} \exp\{2\pi i(ux + vy)\}\, f(e^u,v/\la(-u))
\,\frac{du\,dv}{\la(u)}
\\
&= \iint_{\Aff} \exp\{2\pi i(x\log a + b\la(\log a)y/a)\}\, f(a,b)
\,\frac{da\,db}{a^2}.
\end{align*}
Compare \eqref{eq:quite-dreadful}. The \textit{right Fourier--Moyal
transformation} $\FM^r$ in turn is given by
\begin{align*}
\FM^r[f](z) &:= \int \EE_R(z,g)\, f(g)\,d_rg
\\
&= \iint_{\aff} \exp\{2\pi i(-ux/y + v/y)\} f(e^u,v/\la(-u))
\,\frac{du\,dv}{\la(-u)}
\\
&= \iint_{\Aff} \exp\{2\pi i(-x\log a/y + b\la(\log a)/ay)\}\,
f(a,b)\,\frac{da\,db}{a}.
\end{align*}

We run a first few checks on these Fourier--Moyal kernels. For
$a = 1$, $b = 0$, we recover trivially $\Tr\Om^{L,R}(z) = 1$. Also,
from \eqref{eq:ker-OmU} for $g = 1_{\Aff}$ one gleans without effort
the form \eqref{eq:omz-ker} for the kernel of~$\Om(z)$, that was
useful to invert the Wigner function in subsection~\ref{ssc:Wigner}.

We should be able, as well, to recover the character from the right
kernel, say. Indeed, since
$\int_{\pm\Pi} \Om^R(z) \,d_rz = 1_{\K_\pm}$,
the characters of the representations $U_\pm$ are retrieved from
$$
\int_{\pm\Pi} e^{-2\pi ix\log a/y} \,e^{2\pi ib\la(\log a)/ay}
\,\frac{dx\,dy}{y^2}
= \int_{\R^\x_\pm} \dl(a - 1)\,e^{2\pi iyb} \,\frac{dy}{|y|}
= \chi_\pm(g).
$$
This leads to $\FM^r[\chi_\pm] = \om_\pm$: the ugly duckling of a
character \eqref{eq:Aff-chars} turns here into the swan of the
symplectic form. The same holds for $\FM$. Also, the following
equalities are immediate:
\begin{equation}
\Tr U_\pm[f] = \int_{\pm\Pi} \FM[f](z) \,d_lz
= \int_{\pm\Pi} \FM^r[f](z) \,d_rz.
\label{eq:ratio-regum}
\end{equation}

\subsection{The modified Fourier--Moyal kernels}
\label{ssc:FM-mod}

Just as the operator Fourier transform needs modification for groups
that are not unimodular, we must redefine our Fourier kernels in order
to get a Fourier inversion theorem and a Parseval formula. Consider
now the following distribution or kernel:
$$
\EEmod(z,g) := \Tr\bigl( \Om_\pm(z)\,U_\pm(g)\,d^{1/2}_\pm \bigr)
= W_{U_\pm(g)d^{1/2}_\pm}(z),
$$
for $z \in \pm\Pi$, respectively. Take $\Im z > 0$. From
\eqref{eq:Weyl-Ansatz}, we obtain
$$
\Om(z)\,U(g)\,d^{1/2}
= y \iint_{\R^2} e^{-2\pi i(xu + yv)}\,
U(e^ua, e^u(b + v/\la(u)) M_{\sqrt{\cdot}} \,du\,dv,
$$
and the kernel of this operator is thus
\begin{align*}
\Om Ud^{1/2}(z,g;r,s)
&= y \iint_{\R^2} e^{-2\pi i(xu + yv)}\,
e^{2\pi ie^u(b + v/\la(u))r} \,s^{3/2} \,\dl(s - re^ua) \,du\,dv
\\
&= y \iint_{\R^2} e^{-2\pi i(xu + yv)}\,s^{1/2}\,
e^{2\pi ie^u(b + v/\la(u))r} \,\dl(u - \log(s/ra)) \,du\,dv
\\
&= y \int_\R e^{-2\pi i(x\log(s/ra) + yv)}\,
e^{2\pi ia^{-1}s(b + v/\la(\log(s/ra)))} \,\sqrt{s} \,dv
\\
&= y\, e^{-2\pi ix\log(s/ra)} \,e^{2\pi ibs/a} \,\sqrt{s}
\,\dl\biggl( y - \frac{s}{a\,\la(\log(s/ra))} \biggr).
\end{align*}
Its trace gives us the desired kernel:
\begin{align*}
\EEmod(z,g) &= \int_0^\infty\,\Om Ud^{1/2}(z,g;r,r) \,\frac{dr}{r}
\\
&= y\,e^{2\pi ix\log a} \int_0^\infty e^{2\pi irb/a}
\,\dl\bigl(y - r/\la(\log a)\bigr) \,\frac{dr}{\sqrt r}
\\
&= y\,e^{2\pi ix\log a} \int_0^\infty e^{2\pi irb/a}\,
\la(\log a) \,\dl\bigl(r - y\la(\log a)\bigr) \,\frac{dr}{\sqrt r}
\\
&= \sqrt{|y|\la(\log a)}\,e^{2\pi i(x\log a + y\,b\la(\log a)/a)}.
\end{align*}
We have written $|y|$ for~$y$, so the formula remains valid on the
second orbit. We then see that $\EEmod$ is no longer of modulus~$1$.
We check the coadjoint variation of this kernel, and find that
$$
\EEmod(z,g) = \Dl^{1/2}(h)\,\EEmod(h\lt z,hgh^{-1}).
$$

In conclusion, we make the new definition
\begin{align*}
\FMmod[f](z) &:= \int \EEmod(z,g)\, f(g) \,d_lg
\\
&= \sqrt{|y|}\,\iint_{\aff} \exp\{2\pi i(ux + vy)\}\, f(e^u,v/\la(-u))
\,\frac{du\,dv}{\sqrt{\la(u)}}.
\end{align*}
This is seen to coincide with the $\FAFK$ transform of \cite{AliFK}:
compare~\eqref{eq:Four-Fuhr}.

\vspace{6pt}

Summarizing, the situation is as follows: the formulas analogous
to~\eqref{eq:ratio-regum} are true as well for the Fourier--Kirillov
transform, as pointed out in Section~\ref{sec:Kirillov}; that is of
course well known, and happens for the good reason that the character
is concentrated at $a = 1$ (i.e., at the subgroup generated by the
Puk\'anszky subalgebra subordinate to the maximal orbits). On the
other hand, for the inversion and Plancherel theorems to be perfect
analogues of the ordinary case, we also need a modified Fourier--Moyal
transform. This ``fact of life'' reflects the non-unimodularity of the
group. The relevant results are established in the next subsection.

\subsection{The basic theorems of Fourier analysis}
\label{sec:Planch-punch}

\begin{thm}
\label{thm:FM-props}
The left Fourier--Moyal kernel and transformation enjoy the following
properties:
\begin{itemize}
\item
\emph{Hermiticity:} complex conjugation gives
$\owl{\EE(z,g)} = \EE(z,g^{-1})$.
\item
\emph{Covariance:} $\EE(z,g) = \EE(h \lt z, hgh^{-1})$ for all
$h \in \Aff$.
\item
\emph{Character formula:}
$\displaystyle \int_{\pm\Pi} \EE(z,g) \,d\om_\pm(z) = \Tr U_\pm(g)$.
\item
\emph{Convolution theorem:}
$\FM[f * h] = \FM[f] \star \FM[h]$.
\end{itemize}
Analogous properties hold for the right Fourier--Moyal kernel and map.
\end{thm}

\begin{proof}
The first property is obvious from the definition and the 
selfadjointness of~$\Om(z)$. The second and third properties have 
already been established. The fourth is easy: note first that
\begin{equation}
\EE(\cdot,g) \star \EE(\cdot,g') = \EE(\cdot,gg')
\label{eq:star-repn}
\end{equation}
on account of \eqref{eq:rep-symbol} and~\eqref{eq:run-around}.
Therefore,
$$
\FM[f * h]
= \iint_{\Aff\x\Aff} \EE(\cdot,gg') \,f(g) \,h(g') \,d_lg \,d_lg'
= \FM[f] \star \FM[h].
\eqno \qed
$$
\hideqed
\end{proof}

Analogous properties hold for the modified kernel, as follows.

\begin{thm}
\label{thm:FMmod-props}
The modified Fourier--Moyal kernel and transformation enjoy the
following properties:
\begin{itemize}
\item
\emph{Modified hermiticity:}
$\owl{\EEmod(z,g)} = \Dl^{-1/2}(g)\, \EEmod(z,g^{-1})$.
\item
\emph{Modified covariance:}
$\EEmod(z,g) = \Dl^{1/2}(h)\, \EEmod(h \lt z, hgh^{-1})$ for all
$h \in \Aff$.
\item
\emph{Modified convolution:} for all $f \in L^1(\Aff, d_lg)$ and
$h \in L^1 \cap L^2(\Aff, d_lg)$,
$$
\FMmod(f * h) = \FM[f] \star \FMmod[h] 
= \FMmod[f] \star \FM[\Dl^{-1/2}h].
$$
\end{itemize}
\end{thm}

\begin{proof}
The first two are elementary. For the third, first notice that
$$
\FM[f](z)
= \int_{\Aff} \Tr\bigl( \Om(z)\,U_\pm(g) \bigr) \,f(g) \,d_lg
= \Tr\bigl( \Om(z)\,U_\pm(f) \bigr) = W_{U_\pm(f)}(z)
$$
for $z \in \pm\Pi$, when $f \in L^1(\Aff, d_lg)$. On the other hand, when
$h \in L^1 \cap L^2(\Aff, d_lg)$, the relation
$$
U_\pm(h)\,d^{1/2}_\pm = d^{1/2}_\pm U_\pm(\Dl^{-1/2}h)
$$
follows from selfadjointness of~$d$ and the semi-invariance
relation~\eqref{eq:semi-invce}, integrated over the group. More
precisely, this equality on the domain of $d_\pm^{1/2}$ extends to the
operator closures, which are everywhere defined and Hilbert--Schmidt
on~$\K$ \cite{AliFK,DufloM}. Thus, over the orbits~$\pm\Pi$ we obtain
\begin{align*}
\FMmod(f * h) &= W_{U_\pm(f) U_\pm(h) d_\pm^{1/2}}
= W_{U_\pm(f)\vphantom{d_\pm^{1/2}}} \star W_{U_\pm(h) d_\pm^{1/2}}
= \FM[f] \star \FMmod[h]
\\
&= W_{U_\pm(f) d_\pm^{1/2} U_\pm(\Dl^{-1/2}h)}
= W_{U_\pm(f) d_\pm^{1/2}} \star
  W_{U_\pm(\Dl^{-1/2}h)\vphantom{d_\pm^{1/2}}}
= \FMmod[f] \star \FM[\Dl^{-1/2}h].
\tag*{\qed}
\end{align*}
\hideqed
\end{proof}

The modifications make the nontrivial theorems of Fourier analysis
available: namely, the analogue of the Schur orthogonality relations
for compact groups (the $U_\pm$ are discrete-series representations),
the Fourier inversion theorem and the Plancherel--Parseval unitarity
relation. (See~\cite{Dione} for the compact semisimple case and an
application of it.) To state them, we extend the measure
$\om = \om_+ \cup \om_-$ from $\Pi \cup -\Pi$ to $\aff^*$ by declaring
the complement $\R$ to be a nullset.

\begin{thm}
\label{thm:FMmod-nicer}
These additional properties hold for the modified Fourier--Moyal
kernel and transformation:
\begin{itemize}
\item
\emph{Orthogonality:}
$\displaystyle \int_{\Aff} \owl{\EEmod(z,g)} \,\EEmod(w,g) \,d_lg
= |y|\,\dl(z - w)$ for $z,w \in \pm\Pi$.
\item
\emph{Inversion theorem:}
$\displaystyle f(g)
= \int_{\aff^*} \owl{\EEmod(z,g)} \,\FMmod[f](z) \,d\om(z)$
whenever $f \in \D(\Aff)$.
\item
\emph{Plancherel formula:}
$\displaystyle \int_{\aff^*} \bigl| \FMmod[f](z) \bigr|^2 \,d\om(z)
= \int_{\Aff} |f(g)|^2 \,d_lg$ whenever $f \in L^2(\Aff, d_lg)$.
\end{itemize}
\end{thm}

\begin{proof}
For the Plancherel formula, it is enough to show the relation for
$f \in \D(\Aff)$; the extension to $L^2(\Aff, d_lg)$ by unitarity is
immediate.

Take $z = x + iy$, $w = u + iv$ in $\Pi$; the case of $z,w \in -\Pi$
is similar. The orthogonality relation is straightforward:
\begin{align*}
\int_{\Aff} \owl{\EEmod(z,g)} \, \EEmod(w,g) \,d_lg
&= \int_{\Aff} \sqrt{yv}\, \la(\log a)\,
e^{2\pi i((x - u)\log a + b(v - y)\la(\log a)/a)} \,\frac{da\,db}{a^2}
\\
&= \int_0^\infty \sqrt{yv}\, e^{2\pi i(x - u)\log a} \,\dl(y - v) \,
\frac{da}{a} = |y|\,\dl(x - u)\,\dl(y - v).
\end{align*}
The result obviously implies that
$$
\Om(z) = \int_{\Aff} \owl{\EEmod(z,g)}\, U(g)\, d^{1/2} \,d_lg,
$$
where the $\pm$~signs have been omitted.

For the inversion formula, take $f \in \D(\Aff)$; then
\begin{align*}
\int_{\aff^*} &\owl{\EEmod(z,g)} \,\FMmod[f](z) \,d\om(z)
= \int_{\aff^*} \int_{\Aff} \owl{\EEmod(z,g)} \,\EEmod(z,g') \,
f(g') \,d_lg' \,d\om(z)
\\
&= \int_{\Pi\cup-\Pi} |y| \int_{\Aff} \sqrt{\la(\log a)}\,
\sqrt{\la(\log a')}\, e^{-2\pi i(x(\log a - \log a') 
+ y(b\la(\log a)/a - b'\la(\log a')/a'))}
\\
&\qquad\qquad \x f(a',b') \,\frac{da'\,db'}{a'^2} \,\frac{dx\,dy}{y}
\\
&= \int_\R \int_{\Aff}  \la(\log a')\, \dl(a - a')\,
e^{-2\pi iy(b - b')\la(\log a')/a'} \,f(a',b')
\,\frac{da'\,db'}{a'} \,dy
\\
&= \int_\R \int_\R \frac{\la(\log a)}{a}\,
e^{-2\pi iy(b - b')\la(\log a)/a} \,f(a,b') \,db' \,dy = f(a,b).
\end{align*}
We note the agreeable similarity between the Fourier transformation
and cotransformation.

Finally, the Plancherel relation follows directly from the inversion 
theorem. Again we take $f \in \D(\Aff)$, for simplicity: 
$$
\int_{\aff^*} \bigl| \FMmod[f](z) \bigr|^2 \,d\om(z)
= \int_{\aff^*} \int_{\Aff} \owl{f(g)}\,\, \owl{\EEmod(z,g)}\,
\,\FMmod[f](z) \,d_lg \,d\om(z)
= \int_{\Aff} \owl{f(g)}\, f(g) \,d_lg.
$$
We invite the reader to make a direct proof of this; it proceeds along
the lines of the inversion formula, and is even shorter.
\end{proof}

Recall that in the case of a compact semisimple group, the Plancherel
measure $d\om(z)$ is supported on the integral coadjoint orbits
$\Oh_j$, on each of which the formal dimension $d_j$ is a constant. By
taking $\FMmod(z,g) := d_j^{1/2} \FM(z,g)$ when $z \in \Oh_j$, one
recovers the usual aspect of the Plancherel formula for $\|f\|^2$ as a
weighted sum of integrals over these orbits~\cite{Dione}. The role of
the Duflo--Moore operator as a formal dimension operator is
transparent in our context.

Using Moore's concept of reduced character~\cite{Moore}, defined for
all $f \in \D(\Aff)$, one can establish a character property for
$\FMmod$. We forgo this. Finally, the right Fourier--Moyal kernel and
transformation may be modified in the same way, leading to altogether
analogous harmonic analysis properties, \textit{mutatis mutandis}.

\section{Discussion}
\label{sec:Babel}

\subsection{The Fronsdal program and differential equations for the
Fourier--Moyal kernel}

In the terminology of~\cite{Fronsdal}, our $\EE$-function is a
$\star$-representation, that is, it satisfies
equation~\eqref{eq:star-repn}. In view of the first part of
theorem~\ref{thm:FM-props}, we have a \textit{symmetric}
$\star$-representation (here called hermitian) in the sense of that
reference. Such $\star$-representations are intrinsic objects on
coadjoint orbit, introduced by Fronsdal as a (putative) lifting to the
group level of the $\star$-exponentials of~\cite{BayenFFLS}, which
play a fundamental role in the theory of star products. They fulfil
systems of differential equations. Concretely, Fronsdal's generic
proposal for the $\star$-representation kernel is the locally given
$\star$-exponential:
\begin{equation}
\label{eq:starexp}
\EE_F(F,g) = \exp_\star[2\pi iX](F)
:= \sum_{n=0}^\infty \frac{(2\pi iX)^{\star n}(F)}{n!},
\word{if} g = e^X.
\end{equation}
The coefficient $2\pi$ thrown in here is convenient, given our
definitions. Good treatments of the $\star$-exponential are given by
Arnal~\cite{Arnal88} and Gutt~\cite{Gutt88}. One readily sees that
this object satisfies formally the equation of a
$\star$-representation:
\begin{equation}
\EE_F(\.,g) \star \EE_F(\.,g') = \EE_F(\.,gg').
\label{eq:star-repn-F}
\end{equation}

{}From the covariance relation~\eqref{eq:star-repn-F} one derives
ordinary PDE for this type of $\star$-representation kernel, that may
be sufficient to determine it under favourable circumstances.
Substituting $e^{tX}$ for~$g$ and~$g'$, for any $X \in \g$ one obtains
by differentiation of~\eqref{eq:star-repn-F} at the formal level,
\begin{equation}
\label{eq:diff-star-repn}
[X,\EE_F]_\star = [r(X) - l(X)]\EE_F,
\end{equation}
with $l(X)$, $r(X)$ respectively the corresponding left- and
right-invariant vector fields. We proceed now directly on~$\aff$ and
use its standard basis; then $X_1 \equiv x$, $X_2 \equiv y$. Thus
because of the invariance formulae~\eqref{eq:star-comm}, we must have
in our case:
\begin{align}
y \pd{}{y} \EE(x,y;a,b)
&= \bigl[ r(X_1) - l(X_1) \bigr] \EE(x,y;a,b)
= b \pd{}{b} \EE(x,y;a,b),
\nonumber \\
y \pd{}{x} \EE(x,y;a,b)
&= \bigl[ l(X_2) - r(X_2) \bigr] \EE(x,y;a,b)
= (a - 1) \pd{}{b} \EE(x,y;a,b).
\label{eq:the-gist}
\end{align}
This is the fundamental Fronsdal differential system for~$\Aff$.
Direct inspection of the explicit form~\eqref{eq:trial-by-fire} of
our~$\EE$ shows that these equations are indeed fulfilled.  We already
saw in subsection~\ref{sec:Planch-punch} that the analogue
of~\eqref{eq:star-repn-F} is satisfied by our~$\EE$ as well.

Following the Fronsdal program, invariant affine $\star$-quantization
was studied in~\cite{Huynh}; the latter is the oldest work on
quantization based on the affine group of which we are aware.
Equations~\eqref{eq:the-gist} coincide with equations~(2.9)
of~\cite{Huynh}, when allowance is made for a slightly different
definition of the affine group multiplication. Of course, our focus in
this paper is on the tracial property rather than general covariance.
Thus we did obtain a distinguished solution.

\subsection{Relation with the formalism of Ali \emph{et al}}
\label{sec:veamos-veamos}

In \cite{AliEtal,AliFK}, taken in the context of the affine group of
the line, the Wigner functions are indirectly defined as the images of
the map
$$
\HS(\K_+) \oplus \HS(\K_-) \to L^2(\aff^*, d\om_+ \cup d\om_-),
$$
obtained by composition of the inverse Plancherel transformation and
the $\FAFK$-transformation already given in subsection~\ref{ssc:FK}:
$$
W[A] = \FAFK[P^{-1}(A)].
$$
These authors furthermore propose ``formal Wigner operators'' $W(F)$
via the property
$$
\Tr(A\,W(F)) := W[A](F).
$$
In view of the results in subsection~\ref{sec:Planch-punch}, it is
clear that the dequantization $W_A$ in~\eqref{eq:mouse-trap} and in
subsection~\ref{ssc:Wigner} is the same as $W[A]$, and the formal
Wigner operators $W(F)$ are just the Stratonovich--Weyl (de)quantizers
$\Om(F)$, which are not merely formal at all, and have been explicitly
calculated in this paper. It is remarkable that the integral
expression (49) for $W(F)$ in the first reference in~\cite{AliEtal}
makes manifest use of the Duflo--Moore operators, whereas ours does
not; they however coincide.

Nevertheless, the difference between our axiomatic approach and the
treatment based on the Plancherel transform and the \textit{a priori}
Fourier transformation \eqref{eq:woe-to-come} is not moot. Only the
coadjoint orbits have an interpretation as elementary physical systems
(see \cite{Ganymede} in this respect); the coalgebra $\g^*$ by itself
is an empty vessel. Now, the second definition raises the problem of
the eventual indecomposability of~$W[A]$ on the coadjoint orbits; or,
on account of the Kirillov map, on the unirreps. This
indecomposability actually happens~\cite{AliFK}. Moreover the
definition of $\FAFK$ will not do for compact groups; and the
character formula \textit{\`a la Kirillov} is lost with it, anyway. It
seems preferable to accept that the Stratonovich--Weyl quantizer
generally \textit{determines} the correct scalar Fourier transform,
rather than the other way around, and that for a non-unimodular group
there are \textit{four} such pertinent objects: $\FM$, $\FMmod$
and~$\FM^r$, $\FM^{\mathrm{mod},r}$.

In other words, our approach is geared to fit better with Kirillov
theory. It has an obvious drawback, in that no one knows precisely for
which categories of groups and unirreps do quantizers exist (for
non-type-I groups there is no hope whatsoever). We give an
\textit{aper\c{c}u} of the question in the following subsection,
through the story of Stratonovich--Weyl operators so far.

To conclude, note that our Weyl Ansatz for $Q(f)$ can be rewritten in
the form
$$
Q(f) = \int_{\aff} F[f](u,v) \,U_\pm(e^u, ve^u/\la(u)) \,du\,dv,
$$
where $F$ is the \textit{ordinary} Fourier transformation between
functions on $\aff^*$ and on~$\aff$, and to use it, we extend $f$ by
zero on the complement of~$\Pi$. Thus, the quantization prescription
is not unrelated to the proposal of Manchon~\cite{Manchon} for Weyl
quantization of solvable Lie groups ---which however ignores the issue
of supports within coadjoint orbits, needed to establish boundedness
or compactness of the quantized operators.

\subsection{Setting the record straight}
\label{sec:ajuste-de-cuentas}

The concept of Stratonovich--Weyl quantizer was introduced in the late
eighties \cite{Oberon,Miranda} by two of us. In
\cite[Sect.~3.5]{Polaris} we reported that the name had not caught on,
and called them ``Moyal quantizers'' instead. But the concept itself
certainly did catch on, and beyond \cite{BrifM}, which inaugurated a
wealth of applications, we find it in~\cite{AliEtal} under the name
``Wigner operators''. Lest that nomenclature be misread as a priority
claim, it seems wise to revert to form. We still speak here of
Moyal-type quantization for tracial quantization, and of
Fourier--Moyal kernels and transformations.

The main motivation for the early works was to extend the remit of
phase-space Quantum Mechanics. In particular, tracial twisted products
covariant under~$SU(2)$, for dealing with spinning particles, were
developed in full detail, including applications, in~\cite{Miranda}.
There we were elaborating on old work by Stratonovich
\cite{Stratonovich} ---who should thus be credited with introducing
the ``fuzzy sphere''--- and were unaware of another precedent
\cite{Agarwal}. An equivalent version of the
$SU(2)$-Stratonovich--Weyl quantizer, simpler than our original
expression, is given in~\cite{HeissW}.

The Stratonovich--Weyl quantizer appropriate to deal with relativistic
particles~\cite{Ganymede} was developed shortly after
\cite{Oberon,Miranda}. Indeed, the prevalence of the Heisenberg groups
in quantization is an artifact. From the physical viewpoint, the
coadjoint orbit for the 7-dimensional Heisenberg group makes its
appearance as a direct factor of the \textit{splitting
group}~$\widetilde{\Gal}$ of the covering group of the Galilei
group~\cite{CarinenaOS}, that linearizes its multiplier
representations. Thus the restriction of the quantizer for
$\widetilde{\Gal}$ to the flat part of the orbits renders the standard
Moyal quantizer~\cite{Oberon}; for an explicit calculation showing the
multiplier Galilean covariance of the ordinary Moyal framework,
peruse~\cite{PlebanskiPTT}. All this often goes unremarked.

Stratonovich--Weyl quantizers exist for all compact Lie groups. This
was shown in principle in~\cite{Dione} by the time-honoured method of
interpolating between the ``active'' and ``passive'' symbols
associated to semitracial quantizers. Then in the nineties, apparently
unaware of that work, N.~V. Pedersen introduced a similar set of
postulates, and proved the existence of Stratonovich--Weyl quantizers,
for nilpotent Lie groups~\cite{Pedersen}. See also
\cite[Sect.~4.5]{Fuehr} on this matter. Prior to all that, the
Unterbergers~\cite{Unterbergers} had shown by the interpolation method
the existence of a Moyal-type quantization for the discrete-series
representations of $SL(2,\R)$. In this regard, we wish to
mention~\cite{Orietta} as well. (To our knowledge, however, no one has
been able to exhibit explicitly the Stratonovich--Weyl quantizers for
this case.) These older examples and the work of Ali and coworkers
indicate that Stratonovich--Weyl quantizers exist for large classes of
semi-direct product groups. The time seems ripe for a renewed assault
on Moyal-type quantization and scalar group Fourier transforms
covariant under larger classes of solvable and reductive groups.

Fourier--Moyal kernels are arguably even more important than
Stratonovich--Weyl quantizers, because of their crucial role in
harmonic analysis. They seem destined to complete Kirillov theory. For
years, the abstract nature of expansions of functions on Lie groups in
terms of equivalence classes of unitary transformations has been a
source of some dissatisfaction~\cite{Helgason}. However, one still
finds the Plancherel measure usually realized on~$\widehat G$, rather
than on~$\g^*$. For compact group symmetry, we demonstrated
in~\cite{Dione} how the Fourier--Moyal transformation solves the
problem of giving a formulation of harmonic analysis parallel to
standard Fourier analysis. This section and the previous one show a
wider applicability of its method; and, although here we have opted
for concrete proofs, there is a good chance, in view of the
Leptin--Ludwig theorem, that similar results are valid for all
exponential groups. To finish, we should mention that a bit earlier
---see~\cite{ArnalC} and references therein--- a concept of scalar
``adapted Fourier transform'' had been proposed; because of covariance
trouble it actually does not seem to be all that well adapted to the
context.

\section{Spectral triples on the half-plane}
\label{sec:geometry}

We turn at last to noncommutative spectral triples. The upper
half-plane $\Pi$ is a model for the simplest hyperbolic geometry,
living on a Riemannian surface with negative constant scalar
curvature. It may be regarded as a homogeneous space of the group
$SL(2,\R)$, acting by M\"obius transformations
$z \mapsto (az + b)/(cz + d)$ on~$\Pi$.

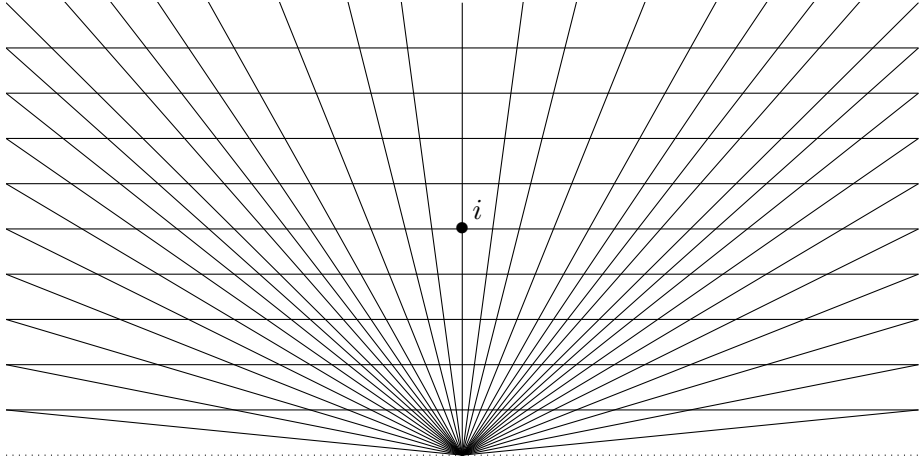
\begin{figure}[htb]
\centering
\begin{tikzpicture}[scale=3]
\clip (-2,0) rectangle (2,2);
\draw[dotted] (-2,0) -- (2,0);
\draw (0,0) -- (0,1) node {$\bullet$} node[above right] {$i$} -- (0,2);
\draw (0,0) -- (2,0.2); \draw (0,0) -- (-2,0.2);
\draw (0,0) -- (2,0.4); \draw (0,0) -- (-2,0.4);
\draw (0,0) -- (2,0.6); \draw (0,0) -- (-2,0.6);
\draw (0,0) -- (2,0.8); \draw (0,0) -- (-2,0.8);
\draw (0,0) -- (2,1.0); \draw (0,0) -- (-2,1.0);
\draw (0,0) -- (2,1.2); \draw (0,0) -- (-2,1.2);
\draw (0,0) -- (2,1.4); \draw (0,0) -- (-2,1.4);
\draw (0,0) -- (2,1.6); \draw (0,0) -- (-2,1.6);
\draw (0,0) -- (2,1.8); \draw (0,0) -- (-2,1.8);
\draw (0,0) -- (2,2.0); \draw (0,0) -- (-2,2.0);
\draw (0,0) -- (2,2.3); \draw (0,0) -- (-2,2.3);
\draw (0,0) -- (2,2.6); \draw (0,0) -- (-2,2.6);
\draw (0,0) -- (2,3.0); \draw (0,0) -- (-2,3.0);
\draw (0,0) -- (2,3.6); \draw (0,0) -- (-2,3.6);
\draw (0,0) -- (2,5.0); \draw (0,0) -- (-2,5.0);
\draw (0,0) -- (2,8.0); \draw (0,0) -- (-2,8.0);
\draw (0,0) -- (2,15.0); \draw (0,0) -- (-2,15.0);
\draw (-2,0.2) -- (2,0.2);
\draw (-2,0.4) -- (2,0.4);
\draw (-2,0.6) -- (2,0.6);
\draw (-2,0.8) -- (2,0.8);
\draw (-2,1.0) -- (2,1.0);
\draw (-2,1.2) -- (2,1.2);
\draw (-2,1.4) -- (2,1.4);
\draw (-2,1.6) -- (2,1.6);
\draw (-2,1.8) -- (2,1.8);
\end{tikzpicture}
\caption{Orbits of $A$ and $N$ for the half-plane}
\label{fg:AN-plane-malla}
\end{figure}

Writing a typical element of the Iwasawa decomposition
$SL(2,\R) = ANK$ as
$$
g = a_t n_s k_\th
= \twobytwo{e^{t/2}}{0}{0}{e^{-t/2}} \twobytwo{1}{s}{0}{1}
\Twobytwo{\cos\half\th}{\sin\half\th}{-\sin\half\th}{\cos\half\th},
$$
the compact subgroup $K = SO(2)$ fixes $i$ and thus
$\Pi \approx SL(2,\R)/SO(2)$ is a principal homogeneous space for the
subgroup $AN$. The orbits of $A$ are half-lines emanating from~$0$,
and the orbits of $N$ are horizontal lines: see
Figure~\ref{fg:AN-plane-malla}.

Under the identification
\begin{equation}
AN \owns a_t n_s \ottto a_t n_s \. i = a_t \.(i + s) = se^t + i e^t
\equiv x + iy = z \in \Pi,
\label{eq:st-coords}
\end{equation}
we may transfer the group operation of $AN$ to~$\Pi$ by letting
$$
(x + iy) \8 (x' + iy') := x + x'y + iyy'.
$$

On the other hand, we may identify $AN$ with $\Aff$ by the group 
isomorphism
\begin{equation}
a_t\,n_s \longmapsto \twobytwo{e^t}{-se^t}{0}{1}.
\label{eq:fearful-decision}
\end{equation}
Observe, however, that the product~$\8$ induced by the (left) M\"obius
action of~$\Aff$ on~$\Pi$ is \textit{opposite} to that
of~\eqref{eq:orbit-prod}, induced by the (left) coadjoint action
of~$\Aff$ on~$\Pi$. This is to be expected, since \eqref{eq:st-coords}
and the identification \eqref{eq:fearful-decision} leads to
$(a,b) = (y,-x)$, which is $g_z^{-1}$ with the
definition~\eqref{eq:root-of-evil}.

The spinor bundle $S \to \Pi$ over the Poincar\'e half-plane has rank
two and is the direct sum of two trivial line bundles, since $\Pi$ is
contractible. Let $\H_0 := L^2(\Pi, y^{-2}\,dx\,dy)$ and let
$\H := \H_0 \oplus \H_0$ be the Hilbert space of spinors. The
Dirac operator is
$\Dslash := -i(\sg_+\,\nb^S_{E_+} + \sg_-\,\nb^S_{E_-})$, using the
isotropic basis $E_+ := 2y\,\del_z = y(\del_x - i\,\del_y)$,
$E_- := 2y\,\delbar_z = y(\del_x + i\,\del_y)$, and the spin
connection $\nb^S_{E_\rho}
:= E_\rho - \quarter \widehat\Ga_{\rho\al}^\bt \,\sg^\al \sg_\bt$ is
determined by the Christoffel symbols of the Levi-Civita connection,
for any zweibein $\{E_\rho\}$. The Levi-Civita connection is canonical
here because~$\Pi$ is a symmetric space~\cite{RieffelPrinc}. Standard
formulas~\cite{Polaris,Nakahara} then yield
$$
\Dslash
= -i \twobytwo{0}{2y\,\del_z + \tihalf}{2y\,\delbar_z - \tihalf}{0}.
$$

Following Palmer \textit{et~al}~\cite{PalmerBT}, one can find
a representation~$\tau$ of $SL(2,\R)$ on~$\H$ under which $\Dslash$
is invariant. It can be written in the form
$$
[\tau(g)\psi](z) := u(g,z) \,\psi(g^{-1} \. z),
$$
where the factor $u$ is of the form
$$
u(g,z) := \twobytwo{v(g^{-1},z)^{1/2}}{0}{0}{v(g^{-1},z)^{-1/2}},
\word{where} v(g,z) := \frac{c\bar z + d}{cz + d}  \word{for}
g = \twobytwo{a}{b}{c}{d},
$$
whenever the square root of $z \mapsto v(g,z)$ can be chosen smoothly,
e.g., for $g$ lying in suitable one-parameter subgroups. A set of
infinitesimal generators $F_0,F_1,F_2$ representing $\gsl(2,\R)$ is
found to be
\begin{align*}
F_0 &= -\half(1 + z^2)\,\del_z - \half(1 - \bar z^2)\,\delbar_z
+ \quarter(z - \bar z) \,\sg_3,
\\
F_1 &= -(z\,\del_z + \bar z\,\delbar_z) = -(x\,\del_x + y\,\del_y),
\qquad  F_0 + F_2 = -\del_x.
\end{align*}
The invariance of $\Dslash$ follows directly from the relations
$[F_j,\Dslash] = 0$. For instance, $[F_1,\Dslash]$ vanishes because
$[x\,\del_x + y\,\del_y, y\,\del_x \mp iy\,\del_y] = 0$. One can
observe that the components of~$\Dslash$ are left-invariant
differential operators on~$\Pi$, regarded as a group. This is why they
commute with the fundamental vector fields $F_1$ and $F_0 + F_2$,
which are of course right-invariant~\cite{Kalliope}.

\vspace{6pt}

Now let a suitably chosen algebra of functions on $\Pi$, under the
twisted product, act diagonally on spinors. This defines on $\Pi$ a
noncommutative operator module in the sense of~\cite{Baer}. That is to
say, there is a noncommutative algebra $(\A,\star)$ involutively
represented by bounded operators on a Hilbert space, and a
self-adjoint operator $D$ on the same Hilbert space, unbounded in the
present case, whose domain is preserved by the action of~$\A$.

We shall now show that the basic pre-condition for a spectral
triple holds, to wit, the commutator of $D$ with the twisted
multiplication by elements of~$\A$ is bounded.

The expression~\eqref{eq:left-of-right} may be applied
componentwise when the function $h$ is replaced by the
two-spinor~$\phi$, namely:
$$
R_w\phi(z)
= \begin{pmatrix} \phi_1(z\.w) \\ \phi_2(z\.w) \end{pmatrix}
= \begin{pmatrix} \phi_1(w \8 z) \\ \phi_2(w \8 z) \end{pmatrix}
=: L^\8_w\phi(z),
$$
with an obvious notation. Now, because the factor $u(g,z)$ is trivial
for $g \in AN$, the invariance of $\Dslash$ under all $\tau(g)$ allows
us to conclude that $L^\8_w \Dslash = \Dslash L^\8_w$, and thus
\begin{align*}
\Dslash(f \star \phi)(z) - f \star \Dslash\phi(z)
&= \iint_{\Pi^2} K_\star(i,w,t)
\bigl( \Dslash_z (L^\8_wf(z) \,L^\8_t \phi(z))
- L^\8_wf(z)\,\Dslash_z \,L^\8_t \phi(z) \bigr) \,d_lw\,d_lt
\\
&= \iint_{\Pi^2} K_\star(i,w,t) \,L^\8_w \,\Dslash f(z)
\,L^\8_t\phi(z) \,d_lw \,d_lt = \Dslash f \star \phi(z),
\end{align*}
where the second equality uses the invariance of the Dirac operator
under the (restriction to~$AN$ of) the $SL(2,\R)$-action by M\"obius
transformations on its orbit~$\Pi$. Thus the Leibniz rule is valid for
the action of~$\Dslash$ on the (left-covariant for the coadjoint
action, right-covariant for the M\"obius action) twisted product. In
particular, if $L_\star(f)$ denotes the operator of left twisted
multiplication, then $[\Dslash, L_\star(f)] = L_\star(\Dslash f)$ is
bounded whenever $f$ has a bounded exterior derivative. Moreover,
arguing as in~\cite{Himalia}, we also get Connes' first-order
condition~\cite{ConnesGrav} from the associativity and complex
conjugation properties of the twisted product. Needless to say, from
here to showing or disproving that the Moyal half-plane in our sense
is a noncommutative geometry, there is still some way to travel.

Some reflection on what has been achieved ---and what has not--- is in
order. The half-plane carries a natural symmetry, namely the left
M\"obius action of $SL(2,\R)$. However, in order to preserve the
Leibniz rule, we have been led to an algebra which is right-invariant,
rather than left-invariant, under (the restriction to the subgroup
$AN$ of) that action. To our knowledge, this was first stated
in~\cite{BieliavskyDSR}, which moreover contains a beautiful study of
the differential equations a general covariant trikernel must satisfy.
If one insists on having left-invariance of the algebra under the
M\"obius action, one can bring into play instead the Dirac operator
associated to the right-invariant metric on~$AN$, which is given by
$$
\Daslash = -i \twobytwo{0}{(x - i)\,\del_x + y\,\del_y + \half}
{(x + i)\,\del_x + y\,\del_y + \half}{0}.
$$
Alternatively, one might try to deform the Dirac operator itself.

\section{Outlook}
\label{sec:finale}

We conclude with a brief review of possible ramifications for our work
in this paper.

\begin{itemize}
\item
Harmonic analysis by way of the scalar Fourier kernels can of course
be pursued much further, around standard lines. For instance, the
third assertion in Theorem~\ref{thm:FMmod-props} remains true when $f$
is a bounded measure. In general, one uses the power $d_\pm^{1/p}$ of
the dimension operators for $L^p$-Fourier analysis. Some matters of
rigour ---see the remark at the end of subsection~\ref{ssc:Wigner}---
will be treated separately~\cite{Eris}. More to the point, the role of
the Fourier--Moyal transformation in relating Wildberger's group class
and coadjoint orbit hypergroups~\cite{WildbergerBimod}, arguably in
the spirit of~\cite{Frobenius}, is an appealing subject of research.
See, with regard to Wildberger's theory, the remarks in
\cite[Sect.~6.4]{Kirillov}.

\item
The issue of $SL(2,\R)$ symmetry for a satisfactory star product is
not ended. It should be obvious that $K_\star^R$ can be extended to
define semitracial, $SL(2,\R)$-covariant star products on the
half-plane. That should allow a fresh attack on the determination of
the Stratonovich--Weyl quantizers for this group~\cite{BieliavskyGI}.
A solution would bring the prize of a suitable and strong definition
of noncommutative Riemann surfaces.

\item
There seems to be no obstruction to the generalization of our method
for constructing star products on $AN$-symmetric spaces, on the basis
of the Iwasawa decomposition. For complex groups, this leads
naturally to Manin triples ---see for instance~\cite{CahenGR}.%
\footnote{Patrizia Vitale pointed this out to us.}
\end{itemize}

\subsection*{Acknowledgments}

Most of this work was done at Universidad Complutense of Madrid (UCM),
at a period when the second named author was a staff member there. We
are thankful for helpful discussions to Paolo Aniello, Pierre
Bieliavsky, Alain Connes, Bruno Iochum, Fedele Lizzi, Giuseppe Marmo,
Patrick V\'arilly, Jasson Vindas, and Patrizia Vitale. Special thanks
are due to Andr\'e Unterberger, who shared with us his unpublished
manuscript~\cite{UnterbergerPrivee}, after the first version of our
article was written. VG thanks the UCM for hospitality. JMG-B
acknowledges partial support from CICyT, Spain, through the grant
FIS2005--02309, and is grateful to Universit\`a di Napoli
Fede\-rico~II for warm hospitality. JCV is grateful to UCM and to
Banco Santander for a Visitante Distinguido fellowship, and thanks the
Universidad de Zaragoza and SISSA, Trieste, for friendly hospitality.
Support for JCV from the Universidad de Costa Rica is acknowledged. We
also thank the referee, whose comments helped to improve the final
manuscript.

\end{document}